\documentclass[pra,superscriptaddress,preprint]{revtex4-1}
\usepackage{graphicx}
\usepackage{amsmath,amssymb,bm}
\usepackage{color}
\usepackage{hyperref}
\hypersetup{colorlinks=true,citecolor=blue,linkcolor=blue,urlcolor=blue}
\def\lesssim{\ \raise.3ex\hbox{$<$}\kern-0.8em\lower.7ex\hbox{$\sim$}\ }
\def\gesim{\ \raise.3ex\hbox{$>$}\kern-0.8em\lower.7ex\hbox{$\sim$}\ }
\begin{document}
\title{Isothermal compressibility and effects of multi-body molecular interactions in a strongly interacting ultracold Fermi gas}
\author{Daichi Kagamihara}
\affiliation{Department of Physics, Kindai University, Higashi-Osaka, Osaka 577-8502, Japan}
\author{Ryohei Sato}
\affiliation{Department of Physics, Keio University, 3-14-1 Hiyoshi, Kohoku-ku, Yokohama 223-8522, Japan}
\author{Koki Manabe}
\affiliation{Department of Physics, Keio University, 3-14-1 Hiyoshi, Kohoku-ku, Yokohama 223-8522, Japan}
\author{Hiroyuki Tajima}
\affiliation{Department of Physics, Graduate School of Science, The University of Tokyo, Tokyo 113-0033, Japan}
\author{Yoji Ohashi}
\affiliation{Department of Physics, Keio University, 3-14-1 Hiyoshi, Kohoku-ku, Yokohama 223-8522, Japan}
\date{\today}
\begin{abstract}
We theoretically investigate the isothermal compressibility $\kappa_{T}$ in the normal state of an ultracold Fermi gas. Including pairing fluctuations, as well as preformed-pair formations, within the framework of the self-consistent $T$-matrix approximation (SCTMA), we evaluate the temperature dependence of this thermodynamic quantity over the entire BCS (Bardeen-Cooper-Schrieffer)-BEC (Bose-Einstein condensation) crossover region. While $\kappa_T$ in the weak-coupling BCS regime is dominated by Fermi atoms near the Fermi surface, correlations between tightly bound Cooper-pair molecules are found to play crucial roles in the strong-coupling BEC regime. In the latter region, besides a two-body molecular interaction, a three-body one is shown to sizably affect $\kappa_T$ near the superfluid phase transition temperature. Our results indicate that the strong-coupling BEC regime of an ultracold Fermi gas would provide a unique opportunity to study multi-body correlations between Cooper-pair molecules. 
\end{abstract}
\maketitle
\par
\section{Introduction}
\par
A pairing interaction between fermions and the resulting Cooper-pair formation are essential ingredients in all Fermi superfluids~\cite{Schrieffer:1999aa}. Particularly in $^{40}\mathrm{K}$ and $^6\mathrm{Li}$ Fermi gases~\cite{Regal:2004aa,Zwierlein:2004aa,Kinast:2004aa,Bartenstein:2004ab}, many-body quantum phenomena originating from a strong pairing interaction have attracted much attention~\cite{Ketterle:2008aa,Bloch:2008aa,Chen:2005aa,Giorgini:2008aa,Zwerger:2011vb}, in connection to the BCS (Bardeen-Cooper-Schrieffer)-BEC (Bose-Einstein condensation) crossover phenomenon~\cite{Eagles:1969aa,Leggett:1980aa,Nozieres:1985aa,Sa-de-Melo:1993aa,Haussmann:1993aa,Haussmann:1994aa,Randeria:1995ta,Timmermans:2001aa,Holland:2001aa,Pistolesi:1994aa,Ohashi:2002aa}: in these Fermi atomic gases, a pairing interaction associated with a Feshbach resonance is tunable by adjusting an external magnetic field~\cite{Chin:2010aa}. Using this advantage, one can continuously change the character of a Fermi superfluid, from the weak-coupling BCS-type to BEC of tightly bound molecules, with increasing the interaction strength. In the intermediate coupling regime, normal-state properties are dominated by fluctuating preformed Cooper pairs, where various interesting many-body phenomena have been discussed both experimentally~\cite{Kinast:2005ab,Altmeyer:2007ul,Stewart:2008aa,Luo:2009aa,Gaebler:2010uy,Nascimbene:2010aa,Navon:2010aa,Cao:2011ab,Sanner:2011aa,Sommer:2011aa,Sommer:2011ab,Ku:2012aa,Elliott:2014ab,Sagi:2015vl,Joseph:2015aa,Horikoshi:2017aa,Hoinka:2017aa} and theoretically~\cite{Pieri:2004aa,Bruun:2005aa,Hu:2006ab,Haussmann:2007aa,Fukushima:2007aa,Hu:2008aa,Tsuchiya:2009aa,Chen:2009aa,Iskin:2009td,Watanabe:2010aa,Hu:2010ab,Magierski:2011aa,Tsuchiya:2011aa,Mueller:2011aa,Perali:2011aa,Enss:2011aa,Palestini:2012aa,Enss:2012ac,Kashimura:2012ti,Wlazlowski:2013ab,Tajima:2014aa,Wyk:2016ab,Ota:2017aa,Tajima:2017aa,Kagamihara:2019ab,Kagamihara:2020aa}.
\par
\begin{figure}[t]
\centering
\includegraphics[width=9cm]{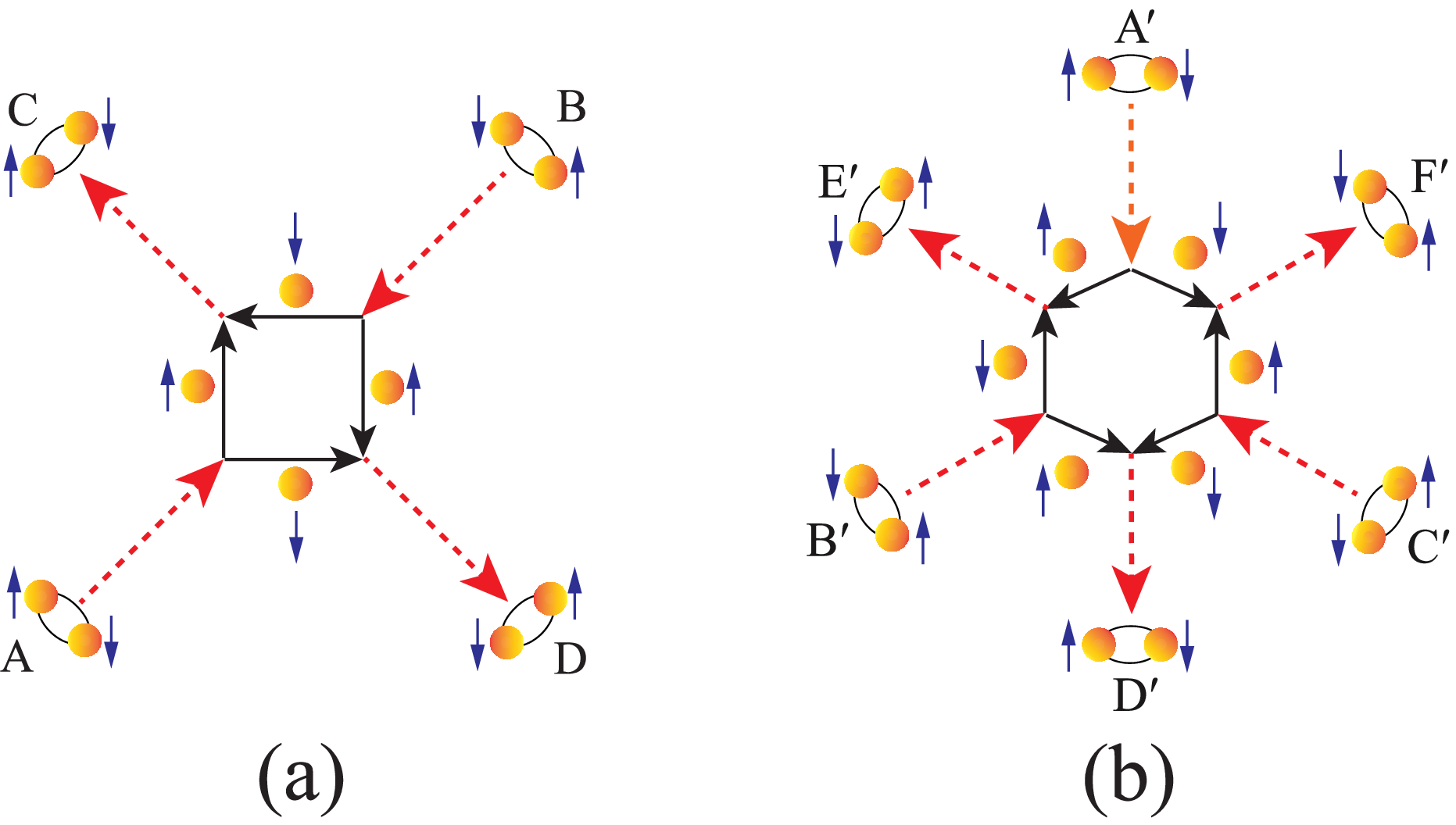}
\caption{Molecular interactions mediated by unpaired Fermi atoms in the strong-coupling BEC regime. (a) Two-body component. (b) Three-body component. The dashed (solid) line describes a bound molecule (dissociated Fermi atom). The pseudo-spin symbols $\sigma=\uparrow,\downarrow$ describe two atomic hyperfine states forming a Cooper pair. In panels (a) and (b), incident molecules (A, B, A$'$, B$'$, C$'$) dissociate into four or fix Fermi atoms, which are followed by recombination to outgoing molecules (C, D, D$'$, E$'$, F$'$).
}
\label{fig1}
\end{figure}

Besides the pairing interaction between fermions, the BCS-BEC crossover phenomenon also provides a unique opportunity to study interactions between Cooper-pair molecules in the strong-coupling BEC regime~\cite{Sa-de-Melo:1993aa,Haussmann:1993aa,Haussmann:1994aa,Pieri:2000aa,Petrov:2004aa,Petrov:2005aa,Ohashi:2005aa,Tempere:2006uu}. Since the ordinary BCS model (which can well describe $^{40}\mathrm{K}$ and $^6\mathrm{Li}$ Fermi gases) only involves an inter-{\it atomic} interaction, such {\it molecular} interactions are mediated by unpaired Fermi atoms, as shown in Fig.~\ref{fig1}(a). Indeed, this diagram is known to give an inter-pair repulsion, being characterized by the $s$-wave molecular scattering length $a_{\mathrm{B}}=2a_s$~\cite{Sa-de-Melo:1993aa,Haussmann:1993aa,Haussmann:1994aa,Engelbrecht:1997wq,Tempere:2006uu}, where $a_s>0$ is the $s$-wave atomic scattering length in the BEC regime. For the value of $a_{\rm B}$, Pieri and Strinati \cite{Pieri:2000aa} pointed out that it becomes small by about the factor three, when multi-scattering processes of the two-body molecular interaction are taken into account. Petrov and co-workers \cite{Petrov:2004aa,Petrov:2005aa} exactly solved a four-fermion problem, to give $a_{\mathrm{B}}\simeq 0.6a_{s}$. Brodsky and co-workers~\cite{Brodsky:2006aa} re-derived this exact molecular scattering length by using a diagrammatic technique. It has also been shown by a renormalization group analysis that many-body corrections lead to a temperature-dependent molecular interaction near the superfluid phase transition temperature $T_{\rm c}$~\cite{Ohashi:2005aa}.  
\par
\par
Although a pairing interaction between fermions is, of course, essentially important in Fermi superfluids, correlations between Cooper pairs also play a crucial role in the superfluid state: in a Bose gas, the superfluid state is known to be unstable against an attractive interaction between bosons~\cite{Pethick:2008aa,Pitaevskii:2016vf}. Thus, the interaction must be repulsive for a Bose superfluid to be stable. In a stable Bose superfluid, the velocity $v_\phi$ of the collective Bogoliubov phonon is directly related to the Bose-Bose repulsion $U_{\rm B}=4\pi a_{\rm B}/M_{\rm B}$ as~\cite{Pethick:2008aa,Pitaevskii:2016vf}
\begin{align}
v_\phi=\sqrt{\frac{U_{\mathrm{B}}N_{\rm c}}{M_{\rm B}}},
\label{eq.0}
\end{align}
where $N_{\rm c}$ is the Bose condensate fraction and $M_{\rm B}$ is a boson mass. (We set $\hbar = k_{\mathrm{B}}=1$ and the system volume $V$ is taken to be unity throughout this paper.) Thus, if an interaction between Cooper-pair `bosons' was attractive, the Fermi superfluid could not stably exist there. The observed sound velocity in the BEC regime of a superfluid $^6\mathrm{Li}$ gas agrees well with Eq.~(\ref{eq.0}) with $a_{\mathrm{B}}=0.6a_s>0$~\cite{Joseph:2007vp}, which means that the molecular interaction is fortunately {\it repulsive} there.
\par
\begin{figure}[t]
\centering
\includegraphics[width=8.6cm]{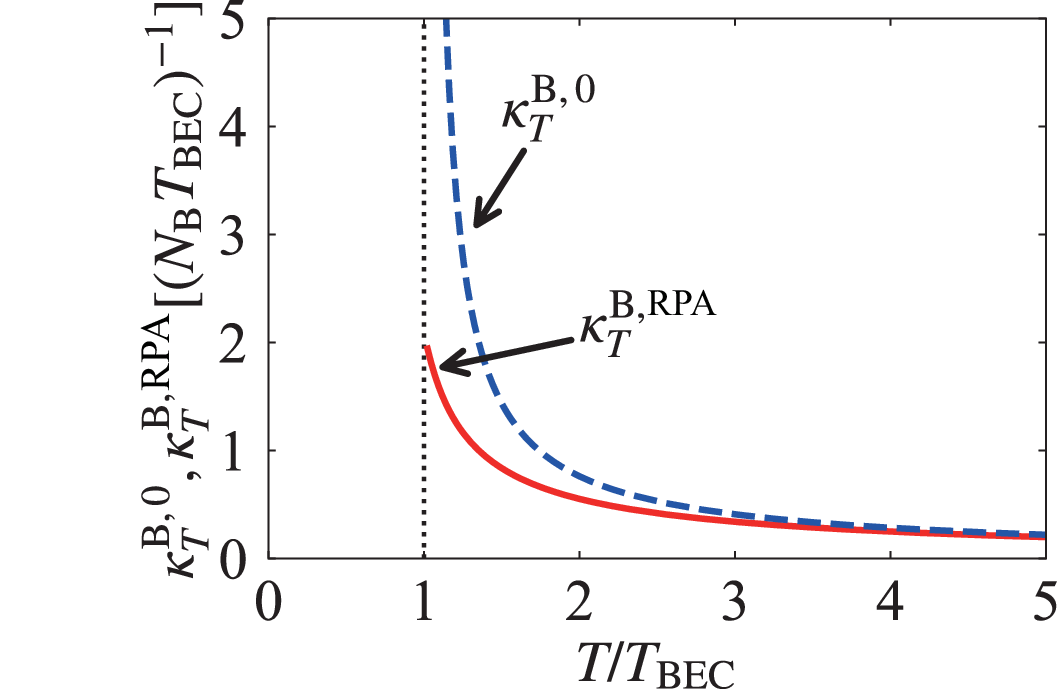}
\caption{Isothermal compressibility in a Bose gas as a function of temperature. $\kappa_T^{\mathrm{B},0}$ and $\kappa_T^{\mathrm{B},\mathrm{RPA}}$, respectively, show the cases of an ideal Bose gas given in Eq.~(\ref{eq.00}) and a repulsively interacting Bose gas given in Eq.~(\ref{eq.2}). In the latter, we take $U_{\mathrm{B}}N_{\mathrm{B}}=0.25T_{\mathrm{BEC}}$, and the Bose-Bose repulsion $U_{\mathrm{B}}$ is treated within the Hartree-Fock-RPA. $N_{\rm B}$ is the number of bosons and $T_{\rm BEC}$ is the BEC phase transition temperature.}
\label{fig2}
\end{figure}
\par
In BCS-BEC crossover physics, the molecular interaction in the BEC regime has so far mainly been discussed within the {\it two-body} level~\cite{Sa-de-Melo:1993aa,Haussmann:1993aa,Haussmann:1994aa,Pieri:2000aa,Petrov:2004aa,Petrov:2005aa,Ohashi:2005aa,Tempere:2006uu}. However, as a simple  extension of Fig.~\ref{fig1}(a), we can also expect, for example, the {\it three-body} molecular interaction illustrated in Fig.~\ref{fig1}(b), which is mediated by six unpaired Fermi atoms. At a glance, since molecular interactions in Fig.~\ref{fig1} are always accompanied by virtual dissociation of molecules in the intermediate state, the resulting interaction seems weaker for a higher-body component. However, the importance of such multi-body molecular interactions is still unclear. Because three-body interactions have also been discussed in various research fields, such as nuclear physics~\cite{Fujita:1957tf,Carlson:2015tw}, as well as neutron-star physics~\cite{Tsubakihara:2013ww}, systematic studies on multi-body molecular correlations by using the high tunability of ultracold Fermi gases would make an impact on these research fields.
\par
The purpose of this paper is to examine how multi-body molecular interactions affect the strong-coupling properties of an ultracold Fermi gas. For this purpose, this paper deals with the isothermal compressibility $\kappa_T$. To explain the reason for this choice, we recall that, as shown in Fig.~\ref{fig2}, the isothermal compressibility in an ideal Bose gas,
\begin{align}
\kappa^{\mathrm{B},0}_T(T)
=&
\frac{1}{N_{\rm B}^2}
\left(
\frac{\partial N_{\rm B}}{\partial \mu_{\rm B}}
\right)_T
\nonumber
\\
=&
\frac{1}{N_{\rm B}^2}
\left(
\frac{\partial}{\partial \mu_{\rm B}}
\sum_{\bm q}
\frac{1}{e^{[\varepsilon_{\bm q}^{\rm B}-\mu_{\rm B}]/T}-1}
\right)_T,
\label{eq.00}
\end{align}
diverges at the Bose-Einstein condensation temperature $T_{\mathrm{BEC}}$ because of $\mu_{\rm B}\to 0$ as
\begin{align}
\kappa_T^{{\rm B},0}(T_{\rm BEC})
=
\frac{1}{N_{\mathrm{B}}^2 T_{\mathrm{BEC}}}
\sum_{\bm q}
{\rm cosech}^2
\left(
\frac{\varepsilon_{\bm q}^{\rm B}}{2T_{\rm BEC}}
\right)
\to \infty.
\label{eq.1}
\end{align}
Here, $\varepsilon_{\bm{q}}^\mathrm{B}=\bm{q}^2/(2M_{\mathrm{B}})$ is the kinetic energy of a boson, $N_{\mathrm{B}}$ the number of bosons, and $\mu_{\mathrm{B}}$ the Bose chemical potential. This divergence at $T_{\rm BEC}$ is absent in the presence of an $s$-wave Bose-Bose repulsion $U_{\mathrm{B}}>0$. Indeed, treating $U_{\mathrm{B}}$ within the Hartree-Fock approximation, one has
\begin{align}
\kappa_T^{\mathrm{B},\mathrm{RPA}}(T)
=&
\frac{1}{N_{\mathrm{B}}^2}
\left(
\frac{\partial}{\partial \mu_{\mathrm{B}}}
\sum_{\bm{q}}
\frac{1}{e^{[\varepsilon_{\bm{q}}^\mathrm{B}+2U_{\rm B}N_{\rm B}-\mu_{\mathrm{B}}]/T}-1}
\right)_T
\nonumber
\\
=&
\kappa_T^{{\rm B},0}(T)
\left[
1 - 2U_{\rm B}
\left(
\frac{\partial N_{\rm B}}{\partial \mu_{\rm B}}
\right)_T
\right].
\label{eq.2b}
\end{align}
Equation~(\ref{eq.2b}) gives the following expression for the isothermal compressibility in the random-phase approximation (RPA):
\begin{align}
\kappa_T^{{\rm B},{\rm RPA}}(T)
=
\frac{\kappa_T^{{\rm B},0}(T)}{1+2U_{\rm B} N_{\rm B}^2\kappa_T^{{\rm B},0}(T)}.
\label{eq.2}
\end{align}
Although the bare isothermal compressibility $\kappa_T^{{\rm B},0}$ diverges at $T_{\rm BEC}$, Eq.~(\ref{eq.2}) converges to give (see also Fig.~\ref{fig2})
\begin{align}
\kappa_T^{{\rm B},{\rm RPA}}(T_{\rm BEC})=\frac{1}{2U_{\rm B}N_{\rm B}^2}.
\label{eq.2z}
\end{align}
This indicates that the isothermal compressibility near the superfluid instability is sensitive to a Bose-Bose interaction. Thus, similar sensitivity is also expected in the BEC regime of an ultracold Fermi gas where most Fermi atoms form tightly bound molecules.
\par
To include strong-coupling effects in the BCS-BEC crossover region, this paper employs the self-consistent $T$-matrix approximation (SCTMA). References~\cite{Haussmann:1993aa,Haussmann:1994aa} showed that the SCTMA gives the molecular scattering length $a_{\rm B}=2a_s$. It has also been shown that the calculated $\kappa_T$ in the SCTMA agrees well with the observed one in a $^6\mathrm{Li}$ unitary Fermi gas~\cite{Sommer:2011aa,Sommer:2011ab,Enss:2012ac}. Thus, the SCTMA is expected to be suitable for our purpose. We briefly note that another well-known BCS-BEC crossover theory called the $T$-matrix approximation (TMA)~\cite{Perali:2002aa} cannot deal with molecular correlations in the normal state~\cite{note1}.
Using the SCTMA scheme, we show that a {\it three-body} molecular interaction sizably affects $\kappa_T$ in the BEC regime. As mentioned previously, although the SCTMA cannot reproduce the exact value of the molecular scattering $a_{\rm B}$~\cite{Petrov:2004aa,Petrov:2005aa}, this strong-coupling scheme is found to still provide useful information about how multi-body molecular interactions work in the BEC regime of an ultracold Fermi gas.
\par
This paper is organized as follows. In Sec.~\ref{section2}, we explain our formulation to evaluate $\kappa_T$ in the SCTMA~\cite{Haussmann:1993aa,Haussmann:1994aa}. We show our results in Sec.~\ref{section3}. In the BEC regime, we evaluate the effects of two-body and three-body molecular interactions from the comparison of our SCTMA result with the isothermal compressibility in an assumed weakly interacting molecular Bose gas. We also compare our result with the recent experiment on a $^6\mathrm{Li}$ unitary Fermi gas~\cite{Ku:2012aa}.
\par
\begin{figure}[t]
\centering
\includegraphics[width=8.6cm]{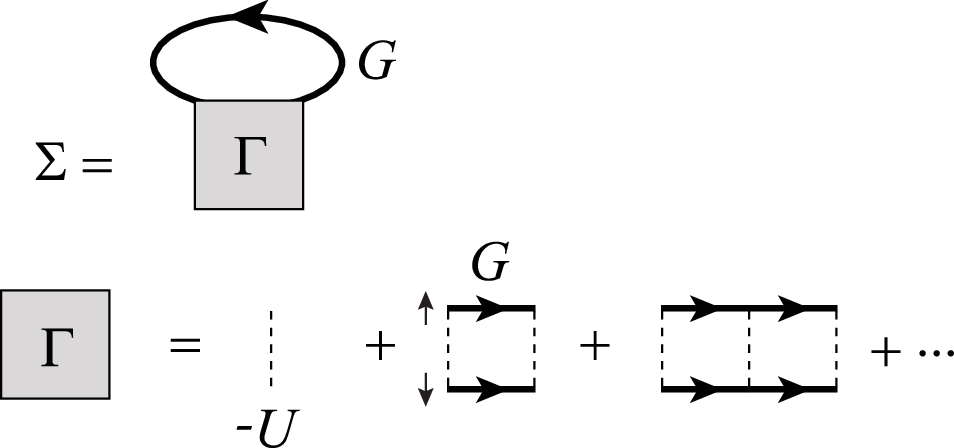}
\caption{SCTMA self-energy correction $\Sigma$. The solid line represents the dressed single-particle thermal Green's function $G$ in Eq.~(\ref{eq.5}). The dotted line is a pairing interaction $-U$. $\Gamma$ is the particle-particle scattering matrix in Eq.~(\ref{eq.8}).
}
\label{fig3}
\end{figure}
\par
\section{Formulation}
\label{section2}
\par
We consider a two-component uniform Fermi gas, described by the BCS Hamiltonian,
\begin{align}
H=\sum_{{\bm p},\sigma}
\xi_{\bm p}c_{{\bm p},\sigma}^\dagger c_{{\bm p},\sigma}
-U
\sum_{\bm{p},\bm{p}',\bm{q}}
c_{{\bm p}+{\bm q}/2,\uparrow}^\dagger
c_{-{\bm p}+{\bm q}/2,\downarrow}^\dagger
c_{-{\bm p}'+{\bm q}/2,\downarrow}
c_{{\bm p}'+{\bm q}/2,\uparrow}.
\label{eq.3}
\end{align}
Here, $c_{{\bm p},\sigma}^\dagger$ is the creation operator of a Fermi atom with pseudo-spin $\sigma=\uparrow,\downarrow$, describing two atomic hyperfine states. The kinetic energy $\xi_{\bm p}=\varepsilon_{\bm p}-\mu={\bm p}^2/(2m)-\mu$ is measured from the Fermi chemical potential $\mu$, where $m$ is an atomic mass. $-U~(<0)$ is a contact-type $s$-wave pairing interaction between Fermi atoms, which is assumed to be tunable by adjusting the threshold energy of a Feshbach resonance~\cite{Chin:2010aa}. We emphasize that Eq.~(\ref{eq.3}) has no term describing any molecular interaction.
\par
We conveniently measure the strength of the pairing interaction in terms of the $s$-wave scattering length $a_s$, which is related to the bare interaction $-U$ as
\begin{align}
\frac{4\pi a_s}{m}=
-
\frac{U}{\displaystyle 1 - U \sum_{\bm p}^{p_{\mathrm{c}}} \frac{1}{2\varepsilon_{\bm p}}},
\label{eq.4}
\end{align}
where $p_{\mathrm{c}}$ is a momentum cutoff. The weak-coupling BCS regime and strong coupling BEC regime are then characterized by $(k_{\mathrm{F}}a_s)^{-1}\lesssim -1$ and $(k_{\mathrm{F}}a_s)^{-1} \gtrsim +1$, respectively (where $k_{\mathrm{F}}$ is the Fermi momentum). The region $-1\lesssim (k_{\mathrm{F}}a_s)^{-1}\lesssim +1$ is sometimes referred to as the (BCS-BEC) crossover region in the literature. 
\par
Strong-coupling corrections to single-particle properties of the system are conveniently described by the self-energy $\Sigma({\bm p},i\omega_n)$ in the dressed Fermi single-particle thermal Green's function, 
\begin{align}
G({\bm p},i\omega_n)=
\frac{1}{G_0({\bm p},i\omega_n)^{-1}-\Sigma({\bm p},i\omega_n)},
\label{eq.5}
\end{align}
where 
\begin{align}
G_0({\bm p},i\omega_n)=
\frac{1}{i\omega_n-\xi_{\bm p}}
\label{eq.6}
\end{align}
is the bare Green's function, with $\omega_n$ being the fermion Matsubara frequency. In the SCTMA, $\Sigma({\bm p},i\omega_n)$ is diagrammatically described as Fig.~\ref{fig3}, which gives
\begin{align}
\Sigma({\bm p},i\omega_n)
=T\sum_{{\bm q},\nu_n}
\Gamma({\bm q},i\nu_n)G({\bm q}-{\bm p},i\nu_n-i\omega_n).
\label{eq.7}
\end{align}
Here, $\nu_n$ is the boson Matsubara frequency and 
\begin{align}
\Gamma({\bm q},i\nu_{n})
=&
-\frac{U}{1-U\Pi({\bm q},i\nu_n)}
\nonumber
\\
=&
\frac{4\pi a_s}{m}
\frac{1}{\displaystyle 1+ \frac{4\pi a_s}{m}
\left[
\Pi({\bm q},i\nu_n)-\sum_{\bm p}\frac{1}{2\varepsilon_{\bm p}}
\right]}
\label{eq.8}
\end{align}
is the SCTMA particle-particle scattering matrix, describing pairing fluctuations. We briefly note that $\Gamma({\bm q},i\nu_n)$ is directly related to a molecular Bose Green's function deep inside the BEC regime~\cite{Haussmann:1993aa,Haussmann:1994aa}. In Eq.~(\ref{eq.8}),
\begin{align}
\Pi({\bm q},i\nu_n)=
T\sum_{\bm{p},\omega_n}
G({\bm p},i\omega_n)
G(\bm{q}-{\bm p},i\nu_n-i\omega_n)
\label{eq.9}
\end{align}
is the pair-correlation function. Although Eq.~(\ref{eq.9}) involves the ultraviolet divergence, it is actually canceled out by the term $\sum_{\bm p}(1/2\varepsilon_{\bm p})$ in Eq.~(\ref{eq.8})~\cite{Haussmann:1993aa,Haussmann:1994aa}. 
\par
\begin{figure}[t]
\centering
\includegraphics[width=10cm]{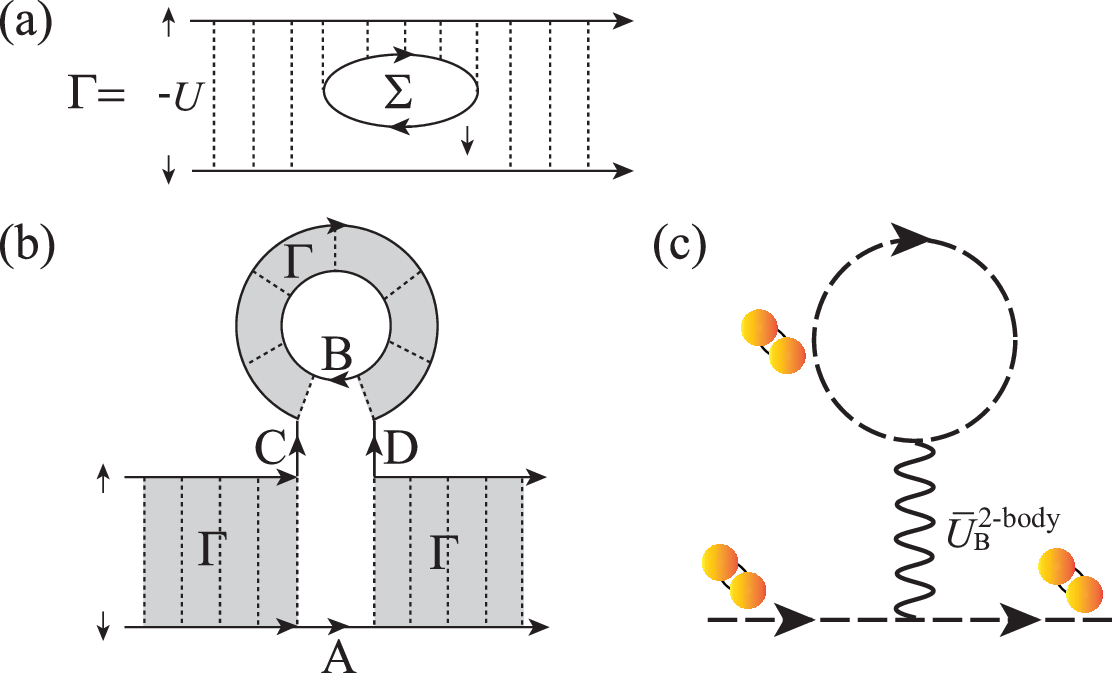}
\caption{Effective two-body molecular interaction involved in the SCTMA~\cite{Haussmann:1993aa,Haussmann:1994aa}. Since the SCTMA particle-particle scattering matrix $\Gamma$ in Eq.~(\ref{eq.8}) consists of the dressed Green's function $G$ (solid line) and the pairing interaction $-U$ (dotted line), $\Gamma$ involves the diagrams shown in panel (a). Without changing the topology, this diagram can be deformed as panel (b). Then, regarding each shaded part in panel (b) as a molecular Bose propagator, one may view the diagram in panel (b) as the Hartree self-energy correction to the molecular Bose Green's function shown in panel (c). In panel (c), the dashed line is the molecular Bose Green's function, and the molecular interaction ${\bar U}_{\mathrm{B}}^{2{\textrm -}{\rm body}}$ corresponds to the four fermion lines A to D in panel (b). The diagrammatic structure giving ${\bar U}_{\mathrm{B}}^{2{\textrm -}{\rm body}}$ is the same as Fig.~\ref{fig1}(a).
}
\label{fig4}
\end{figure}
\par
Here, we explain how the two-body molecular interaction is obtained in the present SCTMA scheme~\cite{Haussmann:1993aa,Haussmann:1994aa}: since $\Gamma$ in Eq.~(\ref{eq.8}) consists of the dressed Green's function $G$, it involves the diagram shown in Fig.~\ref{fig4}(a). Then, deforming this diagram as Fig.~\ref{fig4}(b), and simply regarding the shaded parts as molecular Bose propagators, one finds that Fig.~\ref{fig4}(b) has the same diagrammatic structure as the Hartree self-energy correction to a Bose Green's function shown in Fig.~\ref{fig4}(c). This two-body Bose-Bose interaction ($\equiv {\bar U}_{\mathrm{B}}^{2{\textrm -}{\rm body}}$) is mediated by four unpaired fermions A to D in Fig.~\ref{fig4}(b), and the diagrammatic structure is the same as Fig.~\ref{fig1}(a). References~\cite{Haussmann:1993aa,Haussmann:1994aa} evaluated this molecular interaction in the BEC regime, to give
\begin{align}
{\bar U}_{\rm B}^{2{\textrm -}{\rm body}}=\frac{4\pi (2a_s)}{2m}.
\label{eq.2body}
\end{align}
Recently, the existence of a correction to Eq.~(\ref{eq.2body}) in the SCTMA has been pointed out~\cite{Pini:2019aa}. We will later discuss this from the viewpoint of three-body molecular interaction.
\par
We briefly note that the TMA is obtained by replacing all the dressed Green's function $G$ in Eqs.~(\ref{eq.7}) and (\ref{eq.9}) with the bare one $G_0$ in Eq.~(\ref{eq.6}). The resulting TMA particle-particle scattering matrix does not involve the diagram in Fig.~\ref{fig4}(a). As a result, the non-interacting molecular Bose Green's function is only obtained in this scheme.
\par
\begin{figure}[t]
\centering
\includegraphics[width=6cm]{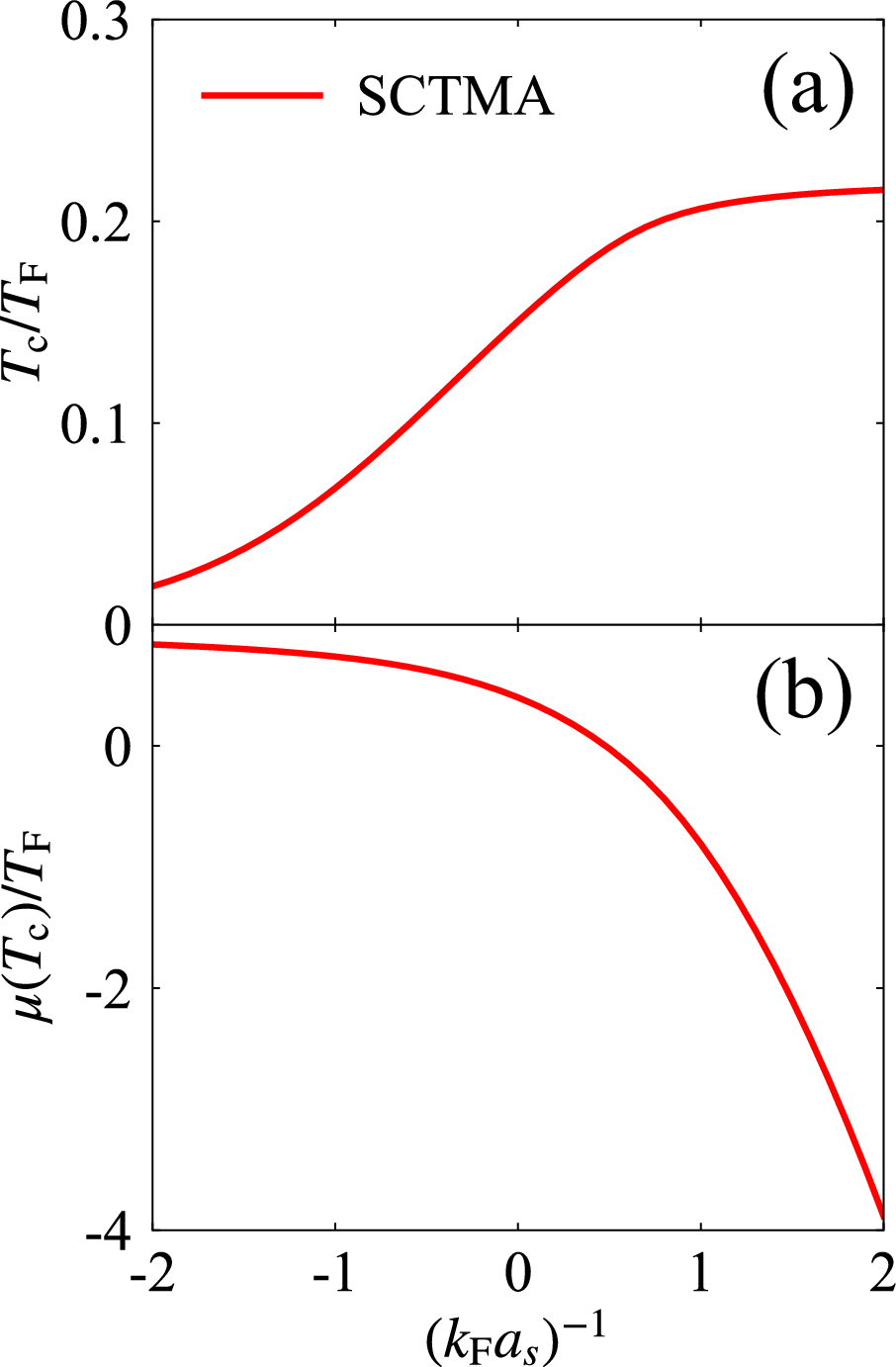}
\caption{SCTMA self-consistent solutions for (a) $T_{\mathrm{c}}$, and (b) $\mu(T_{\mathrm{c}})$. $T_{\rm F}$ is the Fermi temperature.
}
\label{fig5}
\end{figure}

The superfluid phase transition temperature $T_{\rm c}$ is conveniently determined from the Thouless criterion~\cite{Thouless:1960aa}, stating that the system achieves the superfluid instability when the particle-particle scattering matrix $\Gamma$ in Eq.~(\ref{eq.8}) has a pole at ${\bm q}=\nu_{n}=0$, which gives
\begin{align}
1=-\frac{4\pi a_s}{m}
\left[
\Pi(0,0)-\sum_{\bm p} \frac{1}{2\varepsilon_{\bm p}}
\right].
\label{eq.gap}
\end{align}
We actually solve the $T_{\mathrm{c}}$-equation~(\ref{eq.gap}), together with the equation for the total number $N$ of Fermi atoms,
\begin{align}
N=2T\sum_{{\bm p},\omega_{n}}G({\bm p},i\omega_{n}), 
\label{eq.12}
\end{align}
to self-consistently determine $T_{\mathrm{c}}$ and $\mu(T_{\mathrm{c}})$. Above $T_{\mathrm{c}}$, we only deal with Eq.~(\ref{eq.12}) to evaluate $\mu(T>T_{\rm c})$. We briefly show in Fig.~\ref{fig5} the SCTMA solutions for $T_{\rm c}$ and $\mu(T_{\rm c})$, that will be used in evaluating $\kappa_T$. For computational details, see Appendix~\ref{AppendixA}. 
\par
\par
\begin{figure}[t]
\centering
\includegraphics[width=8cm]{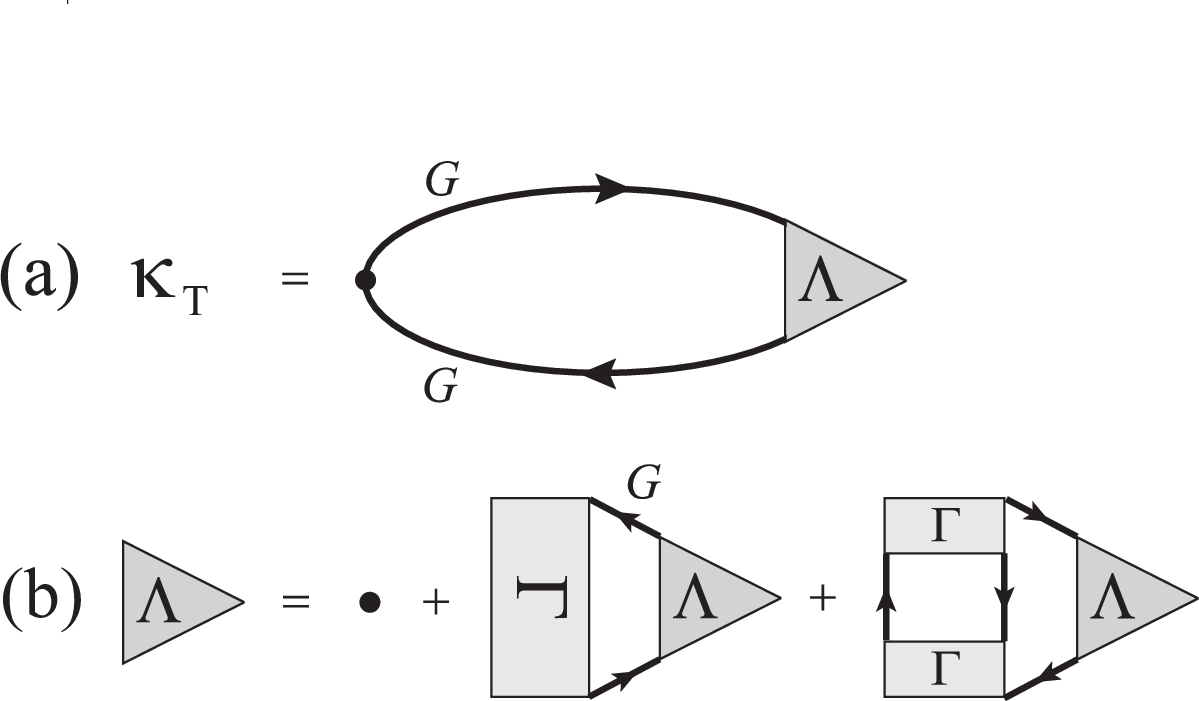}
\caption{(a) Diagrammatic representation of isothermal compressibility $\kappa_T$ in Eq.~(\ref{eq.21b}). The solid circle is the bare density vertex, and $\Lambda$ the three-point vertex correction. (b) Diagrammatic equation for $\Lambda$ in the SCTMA. $\Gamma$ is the particle-particle scattering matrix in Eq.~(\ref{eq.8}). 
}
\label{fig6}
\end{figure}

Once $T_{\rm c}$ and $\mu(T_{\rm c})$, as well as $\mu(T>T_{\rm c})$, are determined, we then evaluate the isothermal compressibility $\kappa_T$ from the following thermodynamic relation:
\begin{align}
\kappa_T=
\frac{1}{N^2}
\left(
\frac{\partial N}{\partial \mu}
\right)_{T}.
\label{eq.13}
\end{align}
 (Note that the system volume is taken to be unity in this paper.) Substituting the number equation~(\ref{eq.12}) into Eq.~(\ref{eq.13}), one obtains
\begin{align}
\kappa_T=
-\frac{2T}{N^2}
\sum_{{\bm p},\omega_n}G^2({\bm p},i\omega_n)\Lambda({\bm p},i\omega_n),
\label{eq.21b}
\end{align}
where the three-point vertex $\Lambda({\bm p},i\omega_n)$ obeys the equation,
\begin{align}
\Lambda({\bm p},i\omega_n)
=&
1-\frac{\partial \Sigma({\bm p},i\omega_n)}{\partial \mu}
\nonumber
\\
=&
1+T
\sum_{{\bm q},\nu_n}
\Gamma({\bm q},i\nu_n)
G^2({\bm q}-{\bm p},i\nu_n-i\omega_n)
\Lambda({\bm q}-{\bm p},i\nu_n-i\omega_n)
\nonumber
\\
&-
2T^2
\sum_{{\bm p}',{\omega}'_n}
\sum_{{\bm q},\nu_n}
G({\bm q}-{\bm p},i\nu_n-i\omega_n)
\Gamma^2({\bm q},i\nu_n)
\nonumber
\\
&~~~~~~~~~~~~~~~~~~~~
\times
G^2({\bm p}',i{\omega}'_n)
G({\bm q}-{\bm p}',i\nu_n-i\omega'_n)
\Lambda({\bm p}',i\omega'_n).
\label{eq.21}
\end{align}
Equations~(\ref{eq.21b}) and (\ref{eq.21}) are diagrammatically described as Figs.~\ref{fig6}(a) and \ref{fig6}(b), respectively. 
\par
When we evaluate $\kappa_T$ from Eq.~(\ref{eq.21b}), we have to self-consistently solve Eq.~(\ref{eq.21}) to determine the vertex correction $\Lambda$. In this paper, to avoid this complicated procedure, we numerically carry out the $\mu$-derivative in Eq.~(\ref{eq.13}) to obtain $\kappa_T$. We will use Eqs.~(\ref{eq.21b}) and (\ref{eq.21}) in Sec.~\ref{subsection3b}, where we examine how molecular interactions affect $\kappa_T$.
\par
\par
\begin{figure}[t]
\centering
\includegraphics[width=8cm]{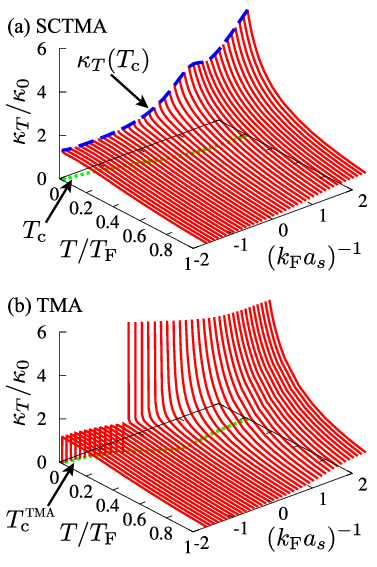}
\caption{Calculated isothermal compressibility $\kappa_T$ in the normal state of an ultracold Fermi gas in the BCS-BEC crossover region. (a) SCTMA. (b) TMA. In panel (b), $\kappa_T$ always diverges at the superfluid phase transition temperature $T_{\mathrm{c}}^{\mathrm{TMA}}$ evaluated in the TMA. (Note that $T_{\mathrm{c}}^{\mathrm{TMA}}$ does not equal $T_{\rm c}$ obtained in the SCTMA.) It diverges positively (negatively) when $(k_{\mathrm{F}} a_s)^{-1} \gtrsim -0.79$ ($(k_{\mathrm{F}} a_s)^{-1} \lesssim -0.79$). $\kappa_0=3m/(k^2_{\mathrm{F}}N)$ is the isothermal compressibility in a free Fermi gas at $T=0$.
}
\label{fig7}
\end{figure}
\par
\section{Isothermal compressibility and effects of molecular interactions}
\label{section3}
\par
\subsection{Isothermal compressibility in the BCS-BEC crossover region}
\par
\begin{figure}[t]
\centering
\includegraphics[width=8cm]{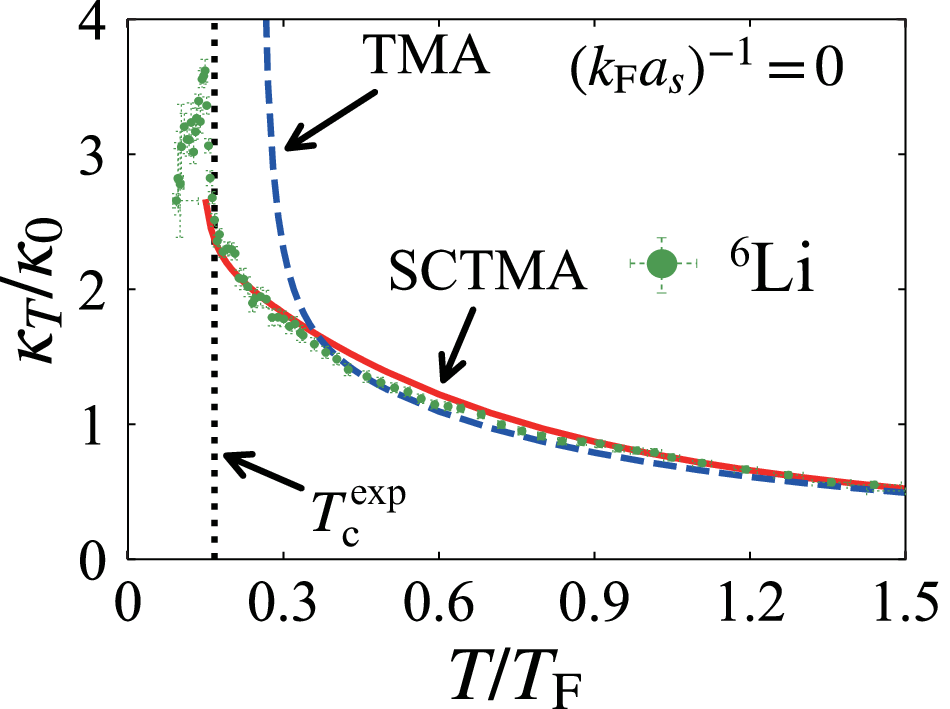}
\caption{Comparison of our theoretical results (SCTMA and TMA) with the recent experiment on a $^6\mathrm{Li}$ unitary Fermi gas~\cite{Ku:2012aa}. $T_{\rm c}^{\rm exp}$ is the superfluid phase transition temperature which is experimentally determined in Ref.~\cite{Ku:2012aa}.
}
\label{fig8}
\end{figure}

Figure~\ref{fig7}(a) shows the SCTMA isothermal compressibility $\kappa_T$ in the normal state of an ultracold Fermi gas in the BCS-BEC crossover region. As expected from the nonzero molecular scattering length $a_{\rm B}=2a_s$~\cite{Haussmann:1993aa,Haussmann:1994aa}, the calculated $\kappa_T$ converges at $T_{\rm c}$ in the whole BCS-BEC crossover region, especially in the BEC regime. This is quite different from the TMA result shown in Fig.~\ref{fig7}(b), where $\kappa_T$ always diverges at $T_{\rm c}$. In the TMA case, the divergence in the BEC regime is due to the ignorance of the molecular interaction. (For more details about the singular behavior seen in Fig.~\ref{fig7}(b), see Appendix \ref{AppendixB}.) Because of this difference, as shown in Fig.~\ref{fig8}, while the SCTMA well explains the experimental result on a $^6\mathrm{Li}$ unitary Fermi gas, the TMA overestimates $\kappa_T$, when $T/T_{\rm F}\lesssim 0.4$ (where $T_{\rm F}$ is the Fermi temperature).
\par
\par
\begin{figure}[t]
\centering
\includegraphics[width=7cm]{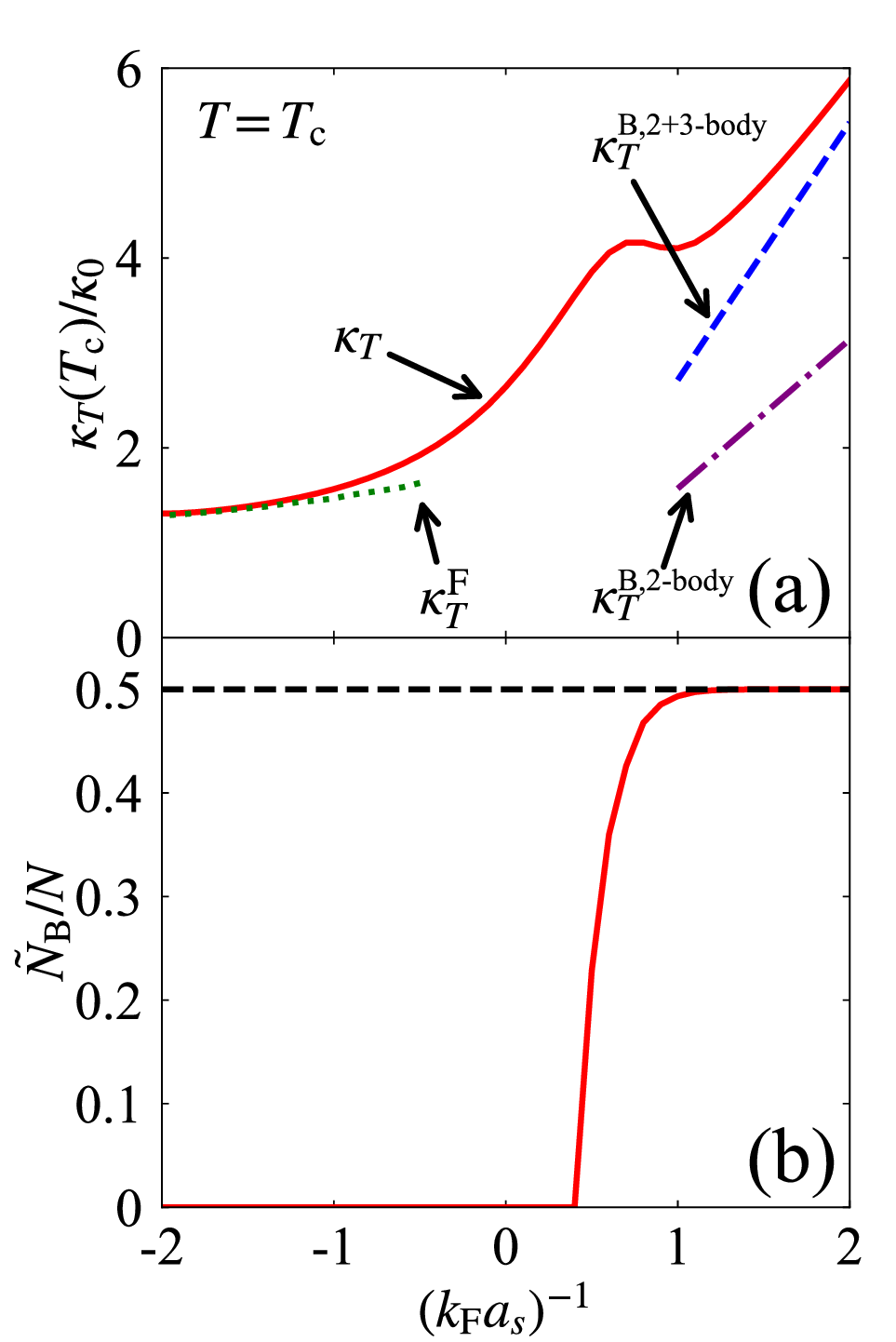}
\caption{(a) SCTMA isothermal compressibility $\kappa_T$ at $T_{\rm c}$. $\kappa_T^{\rm F}$ is the isothermal compressibility in a free Fermi gas given in Eq.~(\ref{eq.33}). $\kappa_T^{{\rm B},2{\textrm-}{\rm body}}$ is the isothermal compressibility in an assumed molecular Bose gas given in Eq.~(\ref{eq.2z}), where $M_{\rm B}=2m$, $N_{\rm B}=N/2$, and $U_{\rm B}={\bar U}_{\rm B}^{2{\textrm -}{\rm body}}=4\pi(2a_s)/M_{\rm B}$. $\kappa_T^{{\rm B},2+3{\textrm -}{\rm body}}$ includes the effects of two-body and three-body molecular interactions, given in Eq.~(\ref{eq.29b}). (b) The number ${\tilde N}_{\rm B}$ of (quasi)stable molecules at $T_{\rm c}$.
}
\label{fig9}
\end{figure}
\par
Figure~\ref{fig9}(a) shows $\kappa_T(T_{\rm c})$ in the SCTMA. In the weak-coupling BCS regime ($(k_{\mathrm{F}}a_s)^{-1} \lesssim -1$), system properties are dominated by Fermi atoms, so that $\kappa_T$ is well described by that in a free Fermi gas,
\begin{align}
\kappa_T^{\mathrm{F}}(T_{\mathrm{c}})
\equiv &
\frac{2}{N^2}
\frac{\partial}{\partial \mu}
\sum_{\bm{p}} \frac{1}{e^{(\varepsilon_{\bm{p}}-{\tilde \mu})/T_{\mathrm{c}}}+1}
\nonumber
\\
=&
\frac{1}{2T_{\mathrm{c}}N^2}
\left( \frac{\partial {\tilde \mu}}{\partial \mu} \right)
\sum_{\bm{p}}
\mathrm{sech}^2 \left( \frac{\varepsilon_{\bm{p}}-{\tilde \mu}}{2T_{\mathrm{c}}} \right).
\label{eq.33}
\end{align}
Here, ${\tilde \mu}={\tilde {\bm k}}_{\rm F}^2/(2m)$ is the effective Fermi chemical potential, where the effective Fermi momentum ${\tilde {\bm k}}_{\rm F}$ is determined from the pole equation of the analytic-continued dressed single-particle Green's function $G(\bm{p},i\omega_n\to\omega+i\delta)$ at $\omega=0$~\cite{Perali:2011aa,Hanai:2014ab},
\begin{align}
\frac{{\tilde {\bm k}}_{\mathrm{F}}^2}{2m}-\mu +
\mathrm{Re}[\Sigma({\tilde {\bm k}}_{\mathrm{F}},i\omega_n\to\omega+i\delta=0+i\delta)]
=0,
\label{eq.32}
\end{align}
with $\delta$ being an infinitesimally small positive number. Because the function ${\rm sech}^2[(\varepsilon_{\bm p}-{\tilde \mu})/2T_{\rm c}]$ in Eq.~(\ref{eq.33}) selectively extracts the contribution around the effective Fermi level $\varepsilon_{\bm p}={\tilde \mu}$, $\kappa_T(T_{\rm c})$ in this regime is dominated by Fermi atoms near this effective Fermi surface.
\par
Starting from the weak-coupling BCS regime, one sees in Fig.~\ref{fig9}(a) that $\kappa_T(T_{\rm c})$ increases with increasing the strength of the pairing interaction. This behavior seen in the BCS side ($(k_{\rm F}a_s)^{-1}\lesssim 0$) is simply due to the well-known property that an attractive Fermi-Fermi interaction enhances the isothermal compressibility.
\par
However, Fig.~\ref{fig9}(a) shows that the increase of $\kappa_T(T_{\rm c})$ is not monotonic, but it exhibits a hump structure around $(k_{\mathrm{F}}a_s)^{-1}=0.7$. Regarding this, estimating the number ${\tilde N}_{\mathrm{B}}$ of (quasi)stable molecules~\cite{note4}, we find in Fig.~\ref{fig9}(b) that ${\tilde N}_{\rm B}$ rapidly increases around $(k_{\mathrm{F}}a_s)^{-1}=0.7$. (We explain how to estimate ${\tilde N}_{\mathrm{B}}$ in Appendix~\ref{AppendixC}.) Thus, the hump structure around $(k_{\mathrm{F}}a_s)^{-1}=0.7$ is considered to be related to the change of dominant particles, from Fermi atoms to Bose molecules~\cite{note3}.
\par
When we simply regard the right region of this hump ($(k_{\rm F}a_s)^{-1} \gtrsim +1$) as a gas of weakly interacting $N/2$ Bose molecules with the {\it two-body} molecular repulsion ${\bar U}_{\rm B}^{2{\textrm -}{\rm body}}$ in Eq.~(\ref{eq.2body}), Eq.~(\ref{eq.2z}) with $U_{\rm B}={\bar U}_{\rm B}^{2{\textrm -}{\rm body}}$ ($\equiv \kappa_T^{{\rm B},2{\textrm -}{\rm body}}$) cannot explain $\kappa_T(T_{\rm c})$, as shown in Fig.~\ref{fig9}(a). In the next subsection, we will show that a three-body molecular interaction resolves this discrepancy.
\par
\par
\subsection{Effects of three-body molecular interaction on $\kappa_T$ in the BEC regime}
\label{subsection3b}
\par
To examine how molecular interactions affect $\kappa_T$ in the BEC regime, we rewrite Eq.~(\ref{eq.21b}) by substituting Eq.~(\ref{eq.21}) into this equation. The resulting expression $\kappa_T=\sum_{j=1}^3\kappa_T^{(j)}$ consists of three terms, where
\begin{align}
\kappa_T^{(1)}
=&
-\frac{2T}{N^2}
\sum_{{\bm p},{\omega}_n}
G^2({\bm p},i{\omega}_n),
\\
\kappa_T^{(2)}
=&
-
\frac{2T^2}{N^2}
\sum_{{\bm p},{\omega}_n}
\sum_{{\bm q},\nu_n}
G^2({\bm p},i{\omega}_n)
\Gamma({\bm q},i\nu_n)
G^2({\bm q}-{\bm p},i\nu_n-i\omega_n)
\Lambda({\bm q}-{\bm p},i\nu_n-i\omega_n),
\\
\kappa_T^{(3)}
=&
\frac{4T^3}{N^2}
\sum_{{\bm p},{\omega}_n}
\sum_{{\bm p}',{\omega}'_n}
\sum_{{\bm q},\nu_n}
G^2({\bm p},i{\omega}_n)
G({\bm q}-{\bm p},i\nu_n-i\omega_n)
\Gamma^2({\bm q},i\nu_n)
\nonumber
\\
\times &
G^2({\bm p}',i{\omega}'_n)
G({\bm q}-{\bm p}',i\nu_n-i\omega'_n)
\Lambda({\bm p}',i\omega'_n).
\label{eq.20}
\end{align}
Among them, $\kappa_T^{(1)}+\kappa_T^{(2)}$ gives the ordinary RPA expression for the isothermal compressibility in a weakly-interacting Fermi gas. Indeed, simply approximating $\Gamma$ by the bare interaction $-U$, and ignoring the last term in Eq.~(\ref{eq.21}), one obtains
\begin{align}
\kappa_T^{(1)}+\kappa_T^{(2)}\simeq 
\frac{\kappa_T^{(1)}}{\displaystyle 1 - \frac{UN^2}{2}\kappa_T^{(1)}}.
\label{eq.100}
\end{align}
To evaluate Eq.~(\ref{eq.100}) in the BEC regime, we recall that, deep inside the BEC regime, the Fermi chemical potential $\mu$ approaches~\cite{Leggett:1980aa,Sa-de-Melo:1993aa,Haussmann:1993aa,Haussmann:1994aa,Randeria:1995ta}
\begin{align}
\mu_{\rm BEC}\equiv-\frac{1}{2ma_s^2}.
\label{eq.BEC}
\end{align}
As a result, $|\mu|$ eventually becomes much larger than the SCTMA self-energy $\Sigma$ involved in the dressed Green's function $G({\bm p},i\omega)$ in Eq.~(\ref{eq.5}). In this case, one may ignore $\Sigma$ compared to $\mu<0$ in evaluating $\kappa_T^{(1)}$, giving
\begin{align}
\kappa_T^{(1)}
\simeq &
-\frac{2T}{N^2}
\sum_{{\bm p},{\omega}_n}
G_{\rm BEC}^2({\bm p},i{\omega}_n)
\nonumber
\\
=&
\frac{1}{2TN^2}
\sum_{\bm p}
{\rm sech}^2 \left( \frac{\varepsilon_{\bm p}+|\mu_{\rm BEC}|}{2T} \right).
\label{eq.kappa_1}
\end{align}
Here, $G_{\rm BEC}({\bm p},i\omega_n)$ has the same form as the bare Green's function $G_0({\bm p},i\omega_n)$ in Eq.~(\ref{eq.6}), where the chemical potential $\mu$ is replaced by $\mu_{\rm BEC}$ in Eq.~(\ref{eq.BEC}). Equation~(\ref{eq.kappa_1}), as well as Eq.~(\ref{eq.100}), vanishes in the BEC limit ($\mu_{\rm BEC}\to-\infty$), so that the isothermal compressibility in the BEC regime is dominated by $\kappa_T^{(3)}$ in Eq.~(\ref{eq.20}).
\par
\begin{figure}[t]
\centering
\includegraphics[width=13cm]{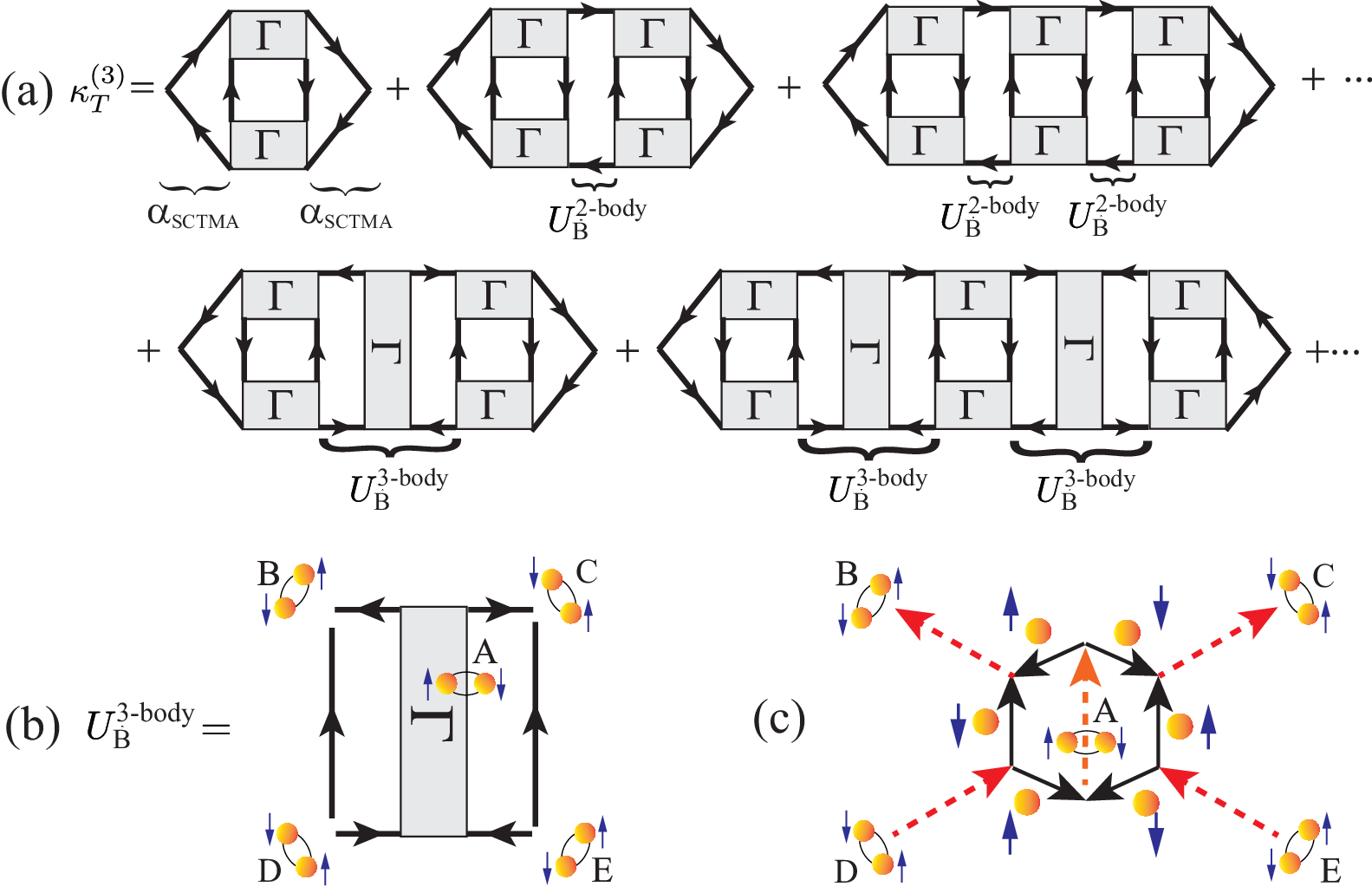}
\caption{(a) Diagrams involved in $\kappa_T^{\rm (3)}$. When $\Gamma$ is regarded as the molecular Green's function, the rectangular part being composed of four Fermi atoms (four solid lines) and the triangular part in the first line may be interpreted as the two-body molecular interaction ($=U_{\rm B}^{2{\textrm -}{\rm body}}$) and the three-point vertex ($=\alpha_{\rm SCTMA}$) of the Bose isothermal compressibility, respectively. The diagrams in the second line involve $U_{\rm B}^{3{\textrm -}{\rm body}}$ shown in panel (b). We show in panel (c) that $U_{\rm B}^{3{\textrm -}{\rm body}}$ is obtained from the three-body molecular interaction in Fig.~\ref{fig1}(b). Molecular propagators A to E in panel (b) correspond to those in panel (c).
}
\label{fig10}
\end{figure}
\par
Repeatedly substituting the three-point vertex $\Lambda$ given in Eq.~(\ref{eq.21}) into Eq.~(\ref{eq.20}), one finds that $\kappa_T^{(3)}$ involves the diagrammatic series shown in Fig.~\ref{fig10}(a). Using the fact that the particle-particle scattering matrix $\Gamma(0,0)$ diverges at $T_{\rm c}$, we approximately evaluate the first term ($\equiv\kappa_{T0}^{(3)}$) in Fig.~\ref{fig10}(a) as
\begin{align}
\kappa_{T0}^{(3)}(T_{\rm c})
\simeq
\frac{\alpha_{\rm SCTMA}^2T_{\rm c}}{N^2}
\sum_{{\bm q},\nu_n}\Gamma^2({\bm q},i\nu_n),
\label{eq.kappa3_0}
\end{align}
where 
\begin{align}
\alpha_{\rm SCTMA}=2T_{\rm c}\sum_{{\bm p},\omega_n}
G^2({\bm p},i\omega_n)G(-{\bm p},-i\omega_n).
\label{eq.vertex}
\end{align}
In the BEC regime, it has been shown that $\Gamma({\bm q},i\nu_n)$ at $T_{\rm c}$ is directly related to the Bose single-particle Green's function $G_{\rm B}^{-1}({\bm q},i\nu_n)=i\nu_n-\varepsilon_{\bm q}^{\rm B}$ as~\cite{Pieri:2000aa,note2}
\begin{align}
\Gamma(\bm{q},i\nu_n)
=Z({\bm q},i\nu)G_{\rm B}({\bm q},i\nu_n).
\label{eq.23}
\end{align}
Here, the molecular mass $M_{\rm B}$ in the Bose kinetic energy $\varepsilon_{\bm q}^{\rm B}={\bm q}^2/2M_{\rm B}$ equals $2m$, and 
\begin{align}
Z(\bm{q},i\nu_n) = \frac{4\pi}{m^2 a_s}
\left[ 1 + 
\sqrt{1 + \frac{-i\nu_n + \varepsilon^{\rm B}_{\bm q}}{E_{\rm bind}}} 
\right],
\label{eq.24}
\end{align}
where $E_{\mathrm{bind}} = 1/(ma_s^2)$ is the binding energy of a two-body bound state. Substitution of Eq.~(\ref{eq.23}) into Eq.~(\ref{eq.kappa3_0}) gives
\begin{align}
\kappa_{T0}^{(3)}(T_{\rm c})
\simeq&
\frac{\alpha_{\rm SCTMA}^2Z^2(0,0)T_{\rm c}}{N^2}
\sum_{{\bm q},\nu_n}G_{\rm  B}^2({\bm q},i\nu_n)
\nonumber
\\
\simeq & 
\frac{1}{4(N/2)^2T_{\rm c}}
\sum_{\bm q}{\rm cosech}^2
\left(\frac{\varepsilon_{\bm q}^{\rm B}}{2T_{\rm c}}\right).
\label{eq.kappa3_BEC}
\end{align}
In obtaining the last expression, we have approximated $G$ involved in $\alpha_{\rm SCTMA}$ to $G_{\rm BEC}$. Equation~(\ref{eq.kappa3_BEC}), which diverges at $T_{\rm c}$, is the same form as Eq.~(\ref{eq.1}) with the boson number $N_{\rm B}=N/2$.
\par
The first line in Fig.~\ref{fig10}(a) is the series of diagrams involving the two-body molecular interaction $U_{\rm B}^{2{\textrm -}{\rm body}}$ mediated by four unpaired fermions. Evaluating these diagrams in the same manner as $\kappa_{T0}^{(3)}$, and adding them to Eq.~(\ref{eq.kappa3_BEC}), they reproduce $\kappa_T^{{\rm B},2{\textrm -}{\rm body}}$ plotted in Fig.~\ref{fig9}(a):
\begin{align}
\kappa_T^{{\rm B},2{\textrm -}{\rm body}}(T_{\rm c})
=
\frac{\kappa_{T0}^{(3)}(T_{\rm c})}{1+2U_{\rm B}^{2{\textrm -}{\rm body}}N_{\rm B}^2\kappa_{T0}^{(3)}(T_{\rm c})}
\to
\frac{1}{2U_{\rm B}^{2{\textrm -}{\rm body}}N_{\rm B}^2}.
\label{eq.29}
\end{align}
Here, 
\begin{align}
U_{\rm B}^{2{\textrm -}{\rm body}}
&=
Z^2(0,0) T \sum_{{\bm p},\omega_n}
G_{\rm BEC}^2({\bm p},i\omega_n)
G_{\rm BEC}^2(-{\bm p},-i\omega_n)
\nonumber
\\
&=
\frac{4\pi (2a_s)}{M_{\rm B}}
\label{eq.30}
\end{align}
just coincides with Eq.~(\ref{eq.2body}). Regarding this, we note that, while ${\bar U}_{\rm B}^{2{\textrm -}{\rm body}}$ in Eq.~(\ref{eq.2body}) is obtained from the Hartree self-energy in Fig.~\ref{fig4}(c), $U_{\rm B}^{2{\textrm -}{\rm body}}$ in Eq.~(\ref{eq.30}) is extracted from the RPA vertex correction to $\kappa_T$. This is a consequence of the present consistent treatment of the SCTMA self-energy $\Sigma$ and the three-point vertex correction $\Lambda$.
\par
The above analysis indicates that the difference between $\kappa_T(T_{\rm c})$ in the SCTMA and $\kappa_T^{{\rm B},2{\textrm -}{\rm body}}(T_{\rm c})$ seen in Fig.~\ref{fig9}(a) comes from the second line in Fig.~\ref{fig10}(a), where each diagram has the part $U_{\rm B}^{3{\textrm -}{\rm body}}$ depicted in Fig.~\ref{fig10}(b). When one again relates $\Gamma$ to the molecular Bose Green's function in the BEC regime, $U_{\rm B}^{3{\textrm -}{\rm body}}$ is found to be obtained from the {\it three-body} molecular interaction given in Fig.~\ref{fig1}(b), as shown in Fig.~\ref{fig10}(c). Because the two of six external lines are contracted in Fig.~\ref{fig10}(c), $U_{\rm B}^{3{\textrm -}{\rm body}}$ may be interpreted as {\it a three-body correction to the two-body molecular interaction}. Evaluating $U_{\rm B}^{3{\textrm -}{\rm body}}$ in the same manner as Eq.~(\ref{eq.30}), one has, in the BEC regime at $T_{\rm c}$,
\begin{align}
U_{\rm B}^{3{\textrm -}{\rm body}}
\simeq&
Z^2(\bm{0},0)
T_{\rm c}^2 \sum_{\bm{p},\bm{p}',\omega_n,\omega_n'}
G_{\rm BEC}^2({\bm p},i\omega_n)
G_{\rm BEC}(-{\bm p},-i\omega_n)
\nonumber
\\
&\times
\Gamma(\bm{p}+\bm{p}',i\omega_n+i\omega_n') G_{\rm BEC}^2(\bm{p}',i\omega_n')
G_{\rm BEC}(-\bm{p}',-i\omega_n')
\nonumber
\\
=&
- \frac{4\pi (0.842a_s)}{M_{\rm B}}. 
\label{eq.31}
\end{align}
For the derivation of Eq.~(\ref{eq.31}), see Appendix~\ref{AppendixD}. 
\par
We comment on the sign of $U_{\rm B}^{3{\textrm -}{\rm body}}$ in Eq.~(\ref{eq.31}): a two-body interaction between molecules is usually considered to be associated with the exchange of constituent Fermi atoms involved in molecules and consequently be repulsive due to the Pauli exclusion principle. Regarding this, Haussmann pointed out that scattering processes contributing to the molecular interaction can be classified into three classes~\cite{Haussmann:1993aa}. The first class is the molecular scattering by a Fermi-Fermi interaction $-U$. The second class is the same as the first one except that the outgoing molecules are exchanged. The third class involves fermion exchange, and this class gives a repulsive interaction due to the Pauli exclusion principle. Haussmann showed in the SCTMA that, while the third class gives Eq.~(\ref{eq.2body})~($\propto a_s)$, the other classes only give corrections in the sub-leading order with respect to $a_s$, by considering the first-order contribution of $-U$. Employing this classification, we find that $U_{\rm B}^{3{\textrm -}{\rm body}}$ in Eq.~(\ref{eq.31}) belongs to the first class, because it is not accompanied by the fermion exchange. Because of this, the sign of $U_{\rm B}^{3{\textrm -}{\rm body}}$ is not attributed to the Pauli exclusion principle, in contrast to $U_{\rm B}^{2{\textrm -}{\rm body}}>0$ in Eq.~(\ref{eq.30}). $U_{\rm B}^{3{\textrm -}{\rm body}}$ becomes $O(a_s)$ due to the multi-scattering processes with respect to $-U$, which is effectively described by the particle-particle scattering matrix $\Gamma$ appearing in the center of Fig.~\ref{fig10}(b), and the resulting sign of this correction becomes negative, as shown in Eq.~(\ref{eq.31}).
\par
Summing up the series of diagrams in both the first and second lines in Fig.~\ref{fig10}(a), as well as diagrams involving both $U_{\rm B}^{2{\textrm -}{\rm body}}$ and $U_{\rm B}^{3{\textrm -}{\rm body}}$ (that are not explicitly shown in Fig.~\ref{fig10}), we reach
\begin{align}
\kappa_T^{{\rm B},2+3{\textrm -}{\rm body}}(T_{\rm c})
=&
\frac{\kappa_{T0}^{(3)}(T_{\rm c})}{1+2[U_{\rm B}^{2{\textrm -}{\rm body}}+U_{\rm B}^{3{\textrm -}{\rm body}}]N_{\rm B}^2\kappa_{T0}^{(3)}(T_{\rm c})}
\nonumber
\\
\to&
\frac{1}{2[U_{\rm B}^{2{\textrm -}{\rm body}}+U_{\rm B}^{3{\textrm -}{\rm body}}]N_{\rm B}^2}.
\label{eq.29b}
\end{align}
Figure~\ref{fig9}(a) shows that this improved result well approaches $\kappa_T$ in the strong-coupling BEC regime ($(k_{\rm F}a_s)^{-1} \gtrsim +1$). This confirms the sizable contribution of the three-body molecular interaction to $\kappa_T$ in this regime. We briefly note that a similar three-body correction has also recently been discussed in a Bose-Fermi mixture~\cite{Manabe:2021vj}.
\par
As mentioned previously, Ref.~\cite{Pini:2019aa} has found from self-energy analyses that the SCTMA scheme actually has a correction to the well-known two-body molecular interaction in Eq.~(\ref{eq.2body}). Our result is consistent with this statement in the sense that this correction is just equal to $U_{\rm B}^{3{\textrm -}{\rm body}}$ in Eq.~(\ref{eq.31}), being obtained from the vertex correction to the isothermal compressibility. This is again a consequence of the consistent treatment of the self-energy $\Sigma$ and the three-point vertex correction $\Lambda$ in our theory. 
\par
The improved molecular scattering length $a_{\rm B}=[2-0.842]a_s=1.158a_s$ obtained from $U_{\rm B}^{2{\textrm -}{\rm body}}+U_{\rm B}^{3{\textrm -}{\rm body}}$ is, however, still larger than the exact value $a_{\rm B}=0.6a_s$~\cite{Petrov:2004aa,Petrov:2005aa}. This means that the SCTMA underestimates $\kappa_T$ in the BEC regime. Since $\kappa_T$ in the SCTMA can explain the recent experiment on a $^6\mathrm{Li}$ unitary Fermi gas~\cite{Ku:2012aa} (see Fig.~\ref{fig8}), the observation of $\kappa_T$ away from the unitary limit would be helpful to see where in the BEC side one needs to improve the SCTMA. 
\par
Here, we compare our result ($a_{\rm B}=1.158a_s$) with the values of $a_{\rm B}$ obtained by various diagrammatic approaches: References~\cite{Haussmann:1993aa,Haussmann:1994aa} examined $a_{\rm B}$ in the SCTMA, to obtain $a_{\rm B} = 2 a_s$. Reference~\cite{Pini:2019aa} studied the self-energy in the SCTMA, to obtain the same result as ours. Reference~\cite{Pieri:2000aa} considered the lowest-order molecular interaction in Eq.~(\ref{eq.2body}), as well as its multi-scattering processes, giving $a_{\rm B} \simeq 0.75 a_s$; however, the three-body correction in Eq.~(\ref{eq.31}) is ignored in this approach. Reference~\cite{Brodsky:2006aa} included all diagrammatic contributions to $a_{\rm B}$, to obtain the exact value $a_{\rm B} \simeq 0.6 a_s$~\cite{Petrov:2004aa,Petrov:2005aa}. Our two-body ($U_{\rm B}^{\operatorname{2-body}}$) and three-body ($U_{\rm B}^{\operatorname{3-body}}$) scattering processes can be seen in Figs.~4(a) and 4(b) in Ref.~\cite{Brodsky:2006aa}, respectively. Although our result ($a_{\rm B}=1.158a_s$) is closer to the exact value $a_{\rm B} \simeq 0.6 a_s$ than $a_{\rm B} = 2 a_s$, to further improve this, we need to include higher order corrections, as well as multi-scattering processes, beyond the SCTMA.
\begin{figure}[t]
\centering
\includegraphics[width=13cm]{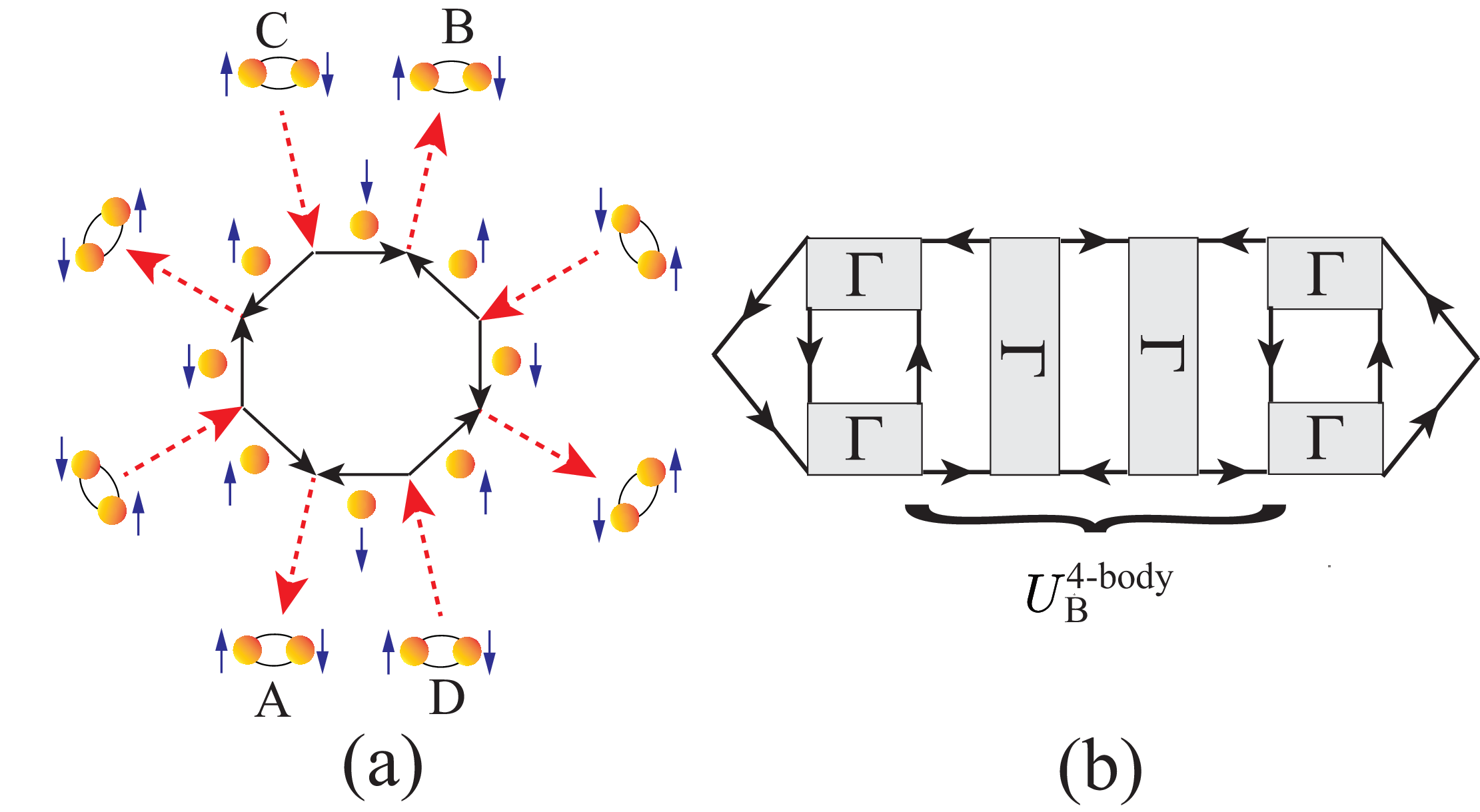}
\caption{(a) Four-body molecular interaction. (b) Diagram involving the effects of the four-body molecular interaction on $\kappa_T$ in the SCTMA. In panel (b), the four-body contribution $U_{\rm B}^{4{\textrm -}{\rm body}}$ is obtained from panel (a) by connecting the two outgoing molecular lines `A' and `B' to the incident lines `C' and `D', respectively. 
}
\label{fig11}
\end{figure}
\par
Before ending this section, we note that, although $\kappa_T^{(3)}$ in Eq.~(\ref{eq.20}) also involves contributions from higher-body molecular interactions $U_{\rm B}^{l{\textrm -}{\rm body}}$ ($l\ge 4$), their contributions are all $O(a_s^n)$ ($n\ge 2$), so that they can be ignored compared to $U_{\rm B}^{2{\textrm -}{\rm body}}\propto a_s$ and $U_{\rm B}^{3{\textrm -}{\rm body}}\propto a_s$, when $(k_{\rm F}a_s)^{-1}\gg 1$. (Although the corresponding diagrams are not shown in Fig. \ref{fig10}(a), for reference, we show in Fig.~\ref{fig11} a four-body molecular interaction and the corresponding diagram involved in $\kappa_T^{(3)}$.) However, it is still unclear whether such higher-body molecular interactions are really irrelevant or their contributions are actually $O(a_s)$ in a more sophisticated strong-coupling theory beyond the SCTMA, which remains as our future problem. Since the effects of multi-body molecular interactions, as well as their multi-scattering processes, should be all taken into account in the exact molecular scattering $a_{\rm B}=0.6a_s$~\cite{Brodsky:2006aa}, in order to assess the importance of higher-body molecular interactions, it would be useful to diagrammatically decompose this exact calculation into contributions from $l$-body molecular interactions. Such analyses might also be helpful in improving the SCTMA so that it can deal with the strong BEC regime in a more quantitative manner.
\par
\par
\section{summary}
\par
To summarize, we have discussed the isothermal compressibility $\kappa_T$ in the BCS-BEC crossover regime of an ultracold Fermi gas above $T_{\rm c}$. Within the framework of the self-consistent $T$-matrix approximation (SCTMA), we have computed $\kappa_T$ in the whole BCS-BEC crossover region. Using the property that this thermodynamic quantity is sensitive to the strength of a Bose-Bose repulsion, we evaluated molecular interactions in the strong-coupling BEC regime.
\par
We showed that $\kappa_T$ monotonically increases with decreasing the temperature in the whole BCS-BEC crossover region, but still converges at $T_{\rm c}$. In the strong-coupling BEC regime where most Fermi atoms form tightly bound molecular bosons, this convergence is attributed to molecular interactions mediated by unpaired Fermi atoms: not only a two-body molecular interaction, but also a three-body one sizably affects this thermodynamic quantity. While the former gives the molecular scattering length $a_{\rm B}=2a_s$ (which is well-known in the SCTMA), the latter corrects this value to $a_{\rm B}=1.158a_s$. This result is consistent with the recent work~\cite{Pini:2019aa}, where the same modified molecular scattering length is obtained from the analysis of the SCTMA self-energy.
\par
As a remaining future problem, although we have clarified the importance of two- and three-body components of molecular interactions in this paper, we still need to examine whether higher-body components are irrelevant or they also sizably affect $\kappa_T$, when one goes beyond the SCTMA. For this problem, it would be helpful to decompose the exact calculation (which gives $a_{\rm B}=0.6a_s$) into the contributions from such multi-body molecular interactions.
\par
In this paper, we have indirectly assessed the effects of the three-body molecular interaction through the correction to the two-body component. Thus, it would also be an interesting future challenge to explore a physical quantity which is more directly affected by multi-body molecular interactions. Since molecular correlations have so far mainly been discussed within the two-body level in ultracold Fermi gases, our results would contribute to the further development of this research field.
\par
\par
\begin{acknowledgments}
We thank M.~Zwierlein for providing us with his experimental data shown in Fig.~\ref{fig8}. We also thank R.~Hanai and D.~Inotani for discussions. D.K. was supported by JST CREST (No.~JPMJCR1673) and JST FOREST (No.~JPMJFR202T). K.M. was supported by a Grant-in-Aid for JSPS fellows (No.~21J14011). H.T. was supported by a Grant-in-Aid for JSPS fellows and Young Scientists (No.~17J03975 and No.~22K13981). Y.O. was supported by a Grant-in-aid for Scientific Research from MEXT and JSPS in Japan (No.~JP18K11345, No.~JP18H05406, No.~JP19K03689, and No.~JP22K03486).
\end{acknowledgments}
\par
\appendix
\par
\section{Computational method to determine \texorpdfstring{$\Sigma({\bm p},i\omega_n)$}{self-energy} in SCTMA}
\label{AppendixA}
\par
\begin{figure}[t]
\centering
\includegraphics[width=10cm]{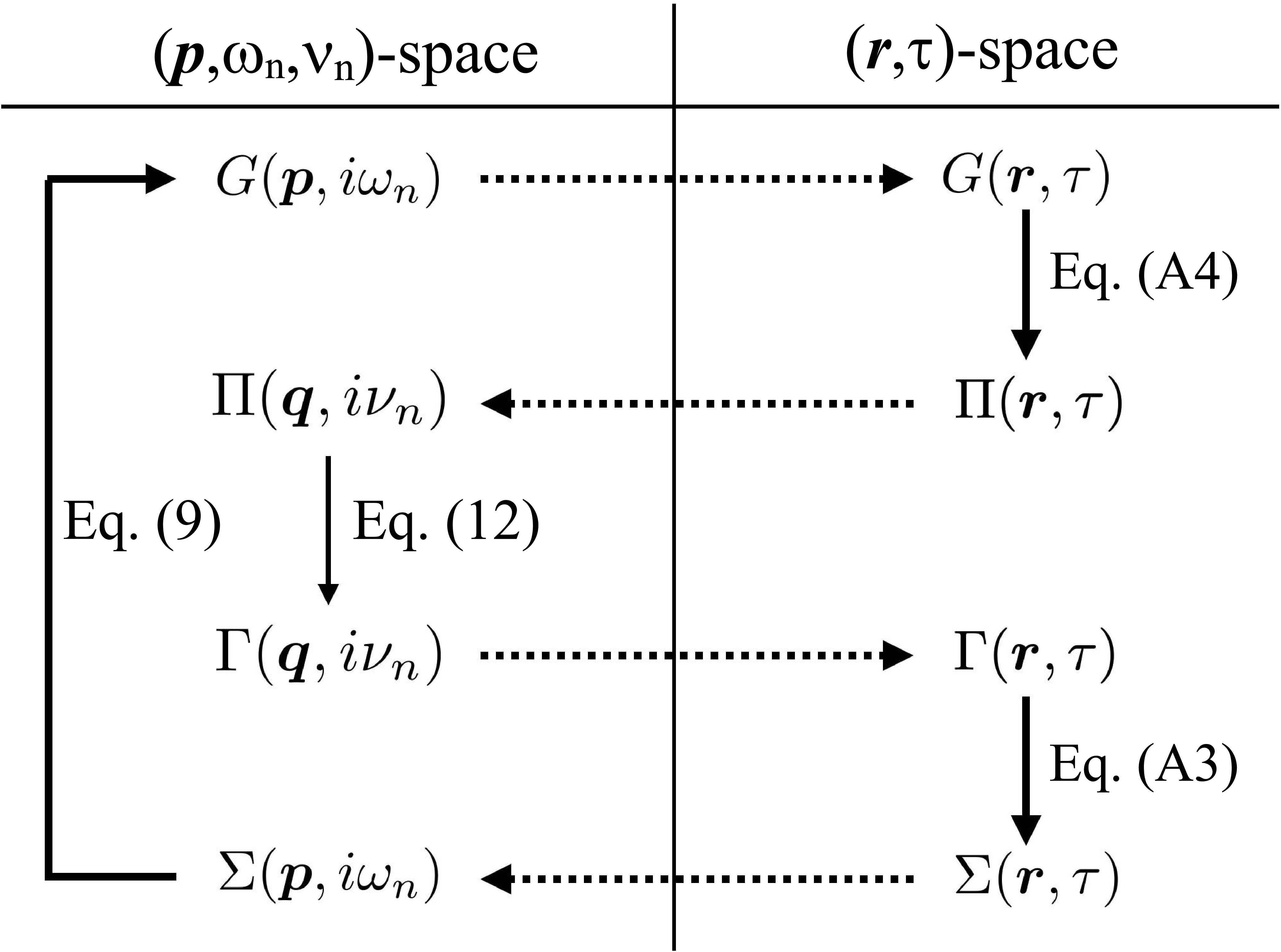}
\caption{Flowchart about self-consistent calculation of SCTMA self-energy  $\Sigma({\bm p},i\omega_n)$ in Eq.~(\ref{eq.7}). The solid arrow means the operation of the equation written beside the arrow. The dashed arrow denotes the Fourier transformation in Eqs.~(\ref{eq.a1}) and (\ref{eq.a2}). We numerically repeat the `calculation loop' in this flowchart, until $\Sigma({\bm p},i\omega_n)$ is self-consistently determined. 
}
\label{fig12}
\end{figure}

In this appendix, we explain how to self-consistently compute the SCTMA self-energy $\Sigma({\bm p},i\omega_n)$ in Eq.~(\ref{eq.7}) from Eqs.~(\ref{eq.5}), (\ref{eq.8}), and (\ref{eq.9}). For this purpose, we make use of the Fourier transform technique~\cite{Haussmann:1994aa,Haussmann:2007aa} in this paper: to avoid computing the momentum and Matsubara-frequency summations in Eqs.~(\ref{eq.7}) and (\ref{eq.9}), we change the variables from `momentum and Matsubara-frequency' to `real space (${\bm r}$) and imaginary time ($\tau$)' by the Fourier transformation,
\begin{align}
f(\bm{r},\tau) &= T \sum_{\bm{p}, \zeta_n} e^{i\bm{p}\cdot \bm{r} - i \zeta_n \tau} f(\bm{p},i\zeta_n),
\label{eq.a1}
\\
f(\bm{p},i\zeta_n) &= \int_0^{1/T} d\tau \int d\bm{r} e^{-i\bm{p}\cdot \bm{r} + i \zeta_n \tau} f(\bm{r},\tau),
\label{eq.a2}
\end{align}
where $\zeta_n$ is the fermion or boson Matsubara frequency. Equations~(\ref{eq.7}) and (\ref{eq.9}) are Fourier-transformed as, respectively,
\begin{align}
\Sigma(\bm{r},\tau) =& \Gamma(\bm{r},\tau) G(-\bm{r},-\tau),
\label{eq.a4}
\end{align}
\begin{align}
\Pi(\bm{r},\tau) =& G(\bm{r},\tau)^2.
\label{eq.a3}
\end{align}
Because Eqs.~(\ref{eq.5}), (\ref{eq.8}), (\ref{eq.a4}), and (\ref{eq.a3}), no longer have any summation, we can quickly compute these. Using this advantage, we self-consistently determine the SCTMA self-energy following the flowchart in Fig.~\ref{fig12}. For the Fourier transformation, we employ the spline interpolation-based Fourier transform technique, developed in Refs.~\cite{Haussmann:1994aa,Haussmann:2007aa}.
\par
\par
\section{Divergence of TMA isothermal compressibility at $T_{\rm c}$}
\label{AppendixB}
\par
The TMA self-energy $\Sigma_{\rm TMA}$ is obtained from Eq.~(\ref{eq.7}) by replacing all the dressed Green's function $G$ with the bare one $G_0$. Evaluating the TMA isothermal compressibility $\kappa_T^{\rm TMA}$ from Eq.~(\ref{eq.13}), we obtain
\begin{align}
\kappa_T^{\rm TMA}
=&
-\frac{2T}{N^2}
\sum_{{\bm p},\omega_n}G_{\rm TMA}^2({\bm p},i\omega_n)
\nonumber
\\
-&
\frac{2T^2}{N^2}
\sum_{{\bm p},\omega_n}\sum_{{\bm q},\nu_n}
G_{\rm TMA}^2({\bm p},i\omega_n)
\Gamma_{\rm TMA}({\bm q},i\nu_n)
G_0^2({\bm q}-{\bm p},i\nu_n-i\omega_n)
\nonumber
\\
+&
\frac{4T^3}{N^2}
\sum_{{\bm p},\omega_n}\sum_{{\bm p}',\omega'_n}\sum_{{\bm q},\nu_n}
G_{\rm TMA}^2({\bm p},i\omega_n)
G_0({\bm q}-{\bm p},i\nu_n-i\omega_n)
\nonumber
\\
\times&
\Gamma_{\rm TMA}^2({\bm q},i\nu_n)
G_0^2({\bm p}',i\omega'_n)
G_0({\bm q}-{\bm p}',i\nu_n-i\omega'_n),
\label{eq.TMA_1}
\end{align}
where the single-particle thermal Green's function $G_{\rm TMA}$ involves the TMA self-energy $\Sigma_{\rm TMA}$ and $\Gamma_{\rm TMA}$ is given by Eq.~(\ref{eq.8}) with $G$ being replaced with $G_0$. At the TMA superfluid phase transition temperature $T_{\rm c}^{\rm TMA}$, the gapless particle-particle scattering matrix behaves as $\Gamma_{\rm TMA}({\bm q},0) \sim 1/q^2$, so that the ${\bm q}$-summation in the last term in Eq.~(\ref{eq.TMA_1}), as well as the resulting $\kappa_T^{\rm TMA}(T_{\rm c}^{\rm TMA})$, always diverges over the entire BCS-BEC crossover region. 
\par
\begin{figure}[t]
\centering
\includegraphics[width=6cm]{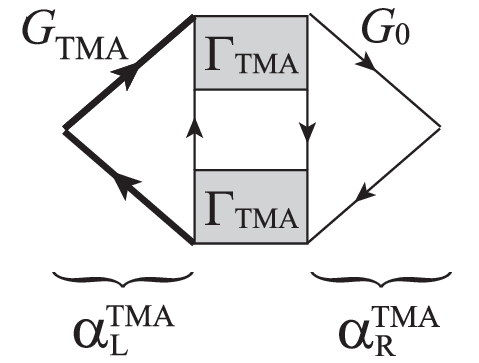}
\caption{Diagrammatic representation of the last term in Eq.~(\ref{eq.TMA_1}). The thick (thin) solid line is the TMA (bare) Green's function. $\alpha_{\rm L}^{\rm TMA}$ and $\alpha_{\rm R}^{\rm TMA}$, respectively, represent the left and right triangular parts in this diagram. $\Gamma_{\rm TMA}$ is the TMA particle-particle scattering matrix. 
}
\label{fig13}
\end{figure}
\par
Using this singular behavior of $\Gamma_{\rm TMA}({\bm q},i\nu_n)$, we only retain the last term in Eq.~(\ref{eq.TMA_1}), to give
\begin{align}
\kappa_T^{\rm TMA}(T_{\rm c}^{\rm TMA})
\simeq
\frac{\alpha_{\rm L}^{\rm TMA} \alpha_{\rm R}^{\rm TMA}}{N^2}
T\sum_{{\bm q},\nu_n}\Gamma_{\rm TMA}^2({\bm q},i\nu_n).
\label{eq.15}
\end{align}
Here, 
\begin{align}
\alpha_{\rm L}^{\rm TMA}=
2T\sum_{{\bm p},\omega_n}
G_{\rm TMA}^2({\bm p},i\omega_n)
G_0(-{\bm p},-i\omega_n),
\label{eq.16}
\end{align}
and
\begin{align}
\alpha_{\rm R}^{\rm TMA}=
2T\sum_{{\bm p},\omega_n}
G_0^2({\bm p},i\omega_n)
G_0(-{\bm p},-i\omega_n),
\label{eq.17}
\end{align}
describe the left and right triangular vertex parts in Fig.~\ref{fig13}, respectively. The product $\alpha_{\rm L}^{\rm TMA} \alpha_{\rm R}^{\rm TMA}$ is numerically found to change its sign at $(k_{\mathrm{F}} a_s)^{-1} \simeq -0.79$, leading to the sign change of $\kappa_{T}^{\rm TMA}(T_{\rm c}^{\rm TMA})$ seen in Fig.~\ref{fig7}(b).
\par
\par
\section{Evaluation of the number $\tilde{N}_{\mathrm{B}}$ of (quasi)stable molecules in Fig.~\ref{fig9}(b)}
\label{AppendixC}
\par
Deep inside the BEC regime ($(k_{\mathrm{F}}a_s)^{-1}\gg 1$), the particle-particle scattering matrix $\Gamma({\bm q},i\nu_n)$ in Eq.~(\ref{eq.8}) is proportional to the single-particle Bose Green's function~\cite{Haussmann:1993aa,Haussmann:1994aa,Pieri:2000aa}. Although this statement is, exactly speaking, only valid for the extreme BEC limit, it is still useful to approximately estimate the number ${\tilde N}_{\mathrm{B}}$ of (quasi)stable molecules in the strong-coupling BEC regime, by assuming a similar relation between $\Gamma({\bm q},i\nu_n)$ and the Bose Green's function given in Eq.~(\ref{eq.23}). That is, ignoring the lifetime of molecules, we determine the molecular excitation energy $\omega_{\bm q}$ from the lowest-energy pole of the analytic continued particle-particle scattering matrix $\Gamma({\bm q},i\nu_n\to\omega_{\bm q}+i\delta)$:
\begin{align}
0=1+\frac{4\pi a_s}{m}
\left[
\mathrm{Re}[\Pi({\bm q},i\nu_n \to \omega_{\bm q}+i\delta)]
-\sum_{\bm p} \frac{1}{2\varepsilon_{\bm{p}}}
\right].
\label{eq.b1}
\end{align}
Here, we have ignored the imaginary part of $\Pi({\bm q},i\nu_n \to \omega_{\bm q}+i\delta)$, for simplicity. 
\par
In the TMA, as well as the strong-coupling theory developed by Nozi\`eres and Schmitt-Rink (NSR)~\cite{Nozieres:1985aa,Ohashi:2002aa,Wyk:2016ab}, the continuum spectrum of $\Pi({\bm q},i\nu_n \to \omega+i\delta)$, which physically describes Fermi excitations being accompanied by the dissociation of molecules, has the clear threshold energy, 
\begin{align}
\omega_{\mathrm{th}}=\frac{\bm{q}^2}{4m}+2|\mu|,
\label{eq.threshold}
\end{align}
in the BEC regime (where $\mu<0$). In this case, we can unambiguously determine the molecular dispersion from the isolated pole below this threshold. 
\par
In contrast, the continuum spectrum does not have such a clear threshold in the SCTMA, because of the self-energy in the dressed Green's function $G$ involved in the pair-correlation function $\Pi$. Thus, in this paper, we approximately employ the threshold energy in Eq.~(\ref{eq.threshold}), and only retain poles below $\omega_{\mathrm{th}}$, in order to distinguish between molecular states and Fermi excitations. Then, simply treating the molecule as a free boson, we estimate the number ${\tilde N}_{\mathrm{B}}$ of (quasi)stable molecules at $T_{\mathrm{c}}$ as
\begin{align}
{\tilde N}_{\mathrm{B}}=\sum_{\bm q}
\frac{1}{e^{\omega_{\bm q}/T_{\mathrm{c}}}-1}.
\label{eq.b2}
\end{align}
\par
We briefly note that the above technique has been used to evaluate $\tilde{N}_{\mathrm{B}}$, as well as the contribution $N_{\rm scatt}$ from the scattering states to the number $N$ of Fermi atoms, in the NSR theory~\cite{Wyk:2016ab}.  Within the NSR scheme, the molecular states are stable with an infinite lifetime.
\par
\par
\section{Derivation of Eq.~(\ref{eq.31})}
\label{AppendixD}
\par
We carry out the Matsubara frequency summations of $\omega_n$ and $\omega_n'$ in Eq.~(\ref{eq.31}), by substituting Eq.~(\ref{eq.23}) into this equation. Approximately setting $e^{\mu/T}=0$ (because $\mu \to -\infty$ in the BEC limit), one has $U_{\rm B}^{3{\textrm -}{\rm body}}=U_{\rm B}^{3{\textrm -}{\rm body}(1)}+U_{\rm B}^{3{\textrm -}{\rm body}(2)}$, where
\begin{align}
U_{\rm B}^{3{\textrm -}{\rm body}(1)} =&
\frac{1}{16}
\left( \frac{8\pi}{m^2a_s}\right)^2
\sum_{\bm{p},\bm{p}'} \frac{1}{\xi_{\bm p}^2\xi_{{\bm p}'}^2} \Gamma(\bm{p}+\bm{p}',-\xi_{\bm{p}}-\xi_{\bm{p}'}),
\label{eq.UB31}
\\
U_{\rm B}^{3{\textrm -}{\rm body}(2)}
=&
\left(\frac{8\pi}{m^2 a_s}\right)^3
\sum_{{\bm p},{\bm q}}
n_{\rm B}(\varepsilon^{\rm B}_{\bm q})
\Biggl[
\frac{1}{\xi_{\bm p}
[\xi_{\bm p}+\xi_{{\bm q}-{\bm p}}-\varepsilon_{\bm q}^{\rm B}]^3
[\xi_{\bm p}-\xi_{{\bm q}-{\bm p}}-\varepsilon_{\bm q}^{\rm B}]
}
\nonumber
\\
{}&~~~~~~~~~~~~~~~~~~~~~
+
\frac{1}{2\xi_{\bm p}
[\xi_{\bm p}+\xi_{{\bm q}-{\bm p}}-\varepsilon_{\bm q}^{\rm B}]^2
[\xi_{\bm p}-\xi_{{\bm q}-{\bm p}}-\varepsilon_{\bm q}^{\rm B}]^2
}
\nonumber
\\
{}&~~~~~~~~~~~~~~~~~~~~~
+\frac{1}{4\xi_{\bm p}^2
[\xi_{\bm p}+\xi_{{\bm q}-{\bm p}}-\varepsilon_{\bm q}^{\rm B}]^2
[\xi_{\bm p}-\xi_{{\bm q}-{\bm p}}-\varepsilon_{\bm q}^{\rm B}]
}
\nonumber
\\
{}&~~~~~~~~~~~~~~~~~~~~~
-\frac{1}{4\xi_{{\bm q}-{\bm p}}^2
[\xi_{\bm p}+\xi_{{\bm q}-{\bm p}}+\varepsilon_{\bm q}^{\rm B}]
[\xi_{\bm p}-\xi_{{\bm q}-{\bm p}}-\varepsilon_{\bm q}^{\rm B}]^2
}
\Biggr].
\label{eq.UB32}
\end{align}
Here, $n_{\rm B}(\varepsilon_{\bm q}^{\rm B}) = [e^{\varepsilon_{\bm q}^{\rm B} / T} - 1]^{-1}$ is the Bose distribution function. 
\par
In Eq.~(\ref{eq.UB31}), we approximately set $\mu\simeq-1/(2ma_s^2)$, as well as change the variables ${\bm p}$, and ${\bm p}'$ as ${\bm p}={\bm k}/a_s$ and ${\bm p}'={\bm k}'/a_s$. Then, we have
\begin{align}
\nonumber
U_{\rm B}^{3{\textrm -}{\rm body}(1)}
=& - \frac{4\pi a_s}{m} \frac{8}{\pi^2}
\int_0^{\infty} k^2dk \int_{0}^{\infty} k'^2 dk' \int_{-1}^{1} d\cos\theta
\\
&~~~~~~~~~~~~~~~~~~~~~~~~~~~
\times
\frac{1}{[k^2 + 1]^2} \frac{1}{[k'^2+1]^2}
\frac{1}{\sqrt{\frac{3}{4}[k^2+k'^2] + \frac{kk'\cos\theta}{2} + 2} - 1},
\label{eq.app10}
\end{align}
where $\theta$ is the angle between $\bm{k}$ and $\bm{k}'$. Numerically evaluating the integrals in Eq.~(\ref{eq.app10}), we obtain
\begin{align}
U_{\rm B}^{3{\textrm -}{\rm body}(1)}
\simeq
 - \frac{4\pi (0.842a_s)}{2m}.
\label{eq.app101B}
\end{align}
\par
For the ${\bm q}$-summation in Eq.~(\ref{eq.UB32}), because the Bose distribution function $n_{\rm B}(\varepsilon_{\bm q}^{\rm B})$ diverges at ${\bm q}=0$, we approximately set ${\bm q}=0$ in this equation except for $\varepsilon_{\bm q}^{\rm B}$ in the Bose distribution function. Again setting $\mu=-1/(2ma_s^2)$, one obtains
\begin{align}
U_{\rm B}^{3{\textrm -}{\rm body}(2)}
\simeq&
- 
\frac{3}{16}\left(\frac{8\pi}{m^2 a_s}\right)^3
\sum_{\bm{q}}
n_{\rm B}(\xi^{\rm B}_{\bm{q}})
\sum_{\bm{p}}
\frac{1}{\xi_{\bm{p}}^5}
=
-\frac{15\pi^2a_s^4N}{m}.
\label{eq.app101}
\end{align}
In obtaining the last expression in Eq.~(\ref{eq.app101}), the molecular number $\sum_{\bm q}n_{\rm B}(\varepsilon_{\bm q}^{\rm B})$ at $T_{\rm c}$ is approximated to half the number $N/2$ of Fermi atoms (because all $N$ Fermi atoms form Bose molecules in the BEC limit). 
\par
While $U_{\rm B}^{3{\textrm -}{\rm body}(1)}=O(a_s)$, $U_{\rm B}^{3{\textrm -}{\rm body}(2)}=O(a_s^4$), so that the former is dominant in the BEC regime. Only retaining the former, we reach
\begin{align}
U_{\rm B}^{3{\textrm -}{\rm body}}
=  - \frac{4\pi (0.842a_s)}{M_{\rm B}}.
\end{align}
where $M_{\mathrm{B}} = 2m$.
\par
\par
\footnotetext[1]{We note that, in contrast to the case of the normal state, the TMA can describe the molecular interaction in the superfluid phase, which comes from the off-diagonal anomalous Green's function. The resulting TMA sound velocity of the Goldstone mode in the BEC regime coincides with the velocity of the Bogoliubov phonon in an assumed Bose superfluid with the molecular scattering length $a_{\mathrm{B}}=2a_s$.}

\footnotetext[2]{We note that the Hartree self-energy in Fig.~\ref{fig4}(c) does not appear in $G_{\rm B}$ at $T_{\rm c}$. This is because it is canceled out by the Bose chemical potential, to give gapless Bose excitations at the superfluid phase transition.}

\footnotetext[3]{We note that, as seen in Fig.~\ref{fig5}(a), the SCTMA does not give the expected hump in $T_{\rm c}$. Regarding this, in a more sophisticated theory which can describe this expected structure, when one moves along this $T_{\rm c}$-line from the weak- to strong-coupling regime, the (quasi)stable molecules should also start to appear in the BEC side, and their effects on $\kappa_T$ would be similar to the SCTMA case. Thus, unless the hump in $\kappa_T$ seen in Fig.~\ref{fig9}(a) has nothing to do with the appearance of molecules, we expect that this structure still remains, even when the hump in $T_{\rm c}$ is taken into account. However, to confirm this expectation, although the TMA~\cite{Perali:2002aa} as well as the strong-coupling theory developed by Nozi\`eres and Schmitt-Rink (NSR)~\cite{Nozieres:1985aa} give a hump in $T_{\rm c}$ around $(k_{\mathrm{F}}a_s)^{-1} \approx 0.4 \sim 0.8$~\cite{Pini:2019aa}, we cannot use these theories, because they unphysically give diverging $\kappa_T(T_{\rm c})$, due to the ignorance of the molecular interaction. It remains as one of our future problems to clarify how the detailed behavior of $T_{\rm c}$ in the BCS-BEC crossover region affects the hump in $\kappa_T$.}

\footnotetext[4]{We note that the number ${\tilde N}_{\mathrm{B}}$ of (quasi)stable molecules does {\it not} involve the contribution ($\equiv N_{\rm scatt}$) from the so-called scattering states~\cite{Nozieres:1985aa,Ohashi:2002aa,Wyk:2016ab}, that physically describe fluctuating Cooper pairs. While the (quasi)stable molecules only appear in the strong-coupling BEC side ($(k_{\rm F}a_s)^{-1} \gesim 0.5$, where $\mu<0$), $N_{\rm scatt}$ first continuously increases with increasing the interaction strength in the BCS regime, reflecting the enhancement of pairing fluctuations; however, $N_{\rm scatt}$ decreases to eventually disappear deep inside the BEC regime where system properties are dominated by (quasi)stable molecules. Since the molecular states obtained in the SCTMA have a small but non-zero decay rate (which is, however, ignored in calculating ${\tilde N}_{\rm B}$, see Appendix~\ref{AppendixC}), we call them ``(quasi)''stable molecules in this paper.}
\par

\bibliography{library}

\begin{thebibliography}{87}
\bibitem [{Note1()}]{note1}%
  \BibitemOpen
  \bibinfo {note} {We note that, in contrast to the case of the normal state,
  the TMA can describe the molecular interaction in the superfluid phase, which
  comes from the off-diagonal anomalous Green's function. The resulting TMA
  sound velocity of the Goldstone mode in the BEC regime coincides with the
  velocity of the Bogoliubov phonon in an assumed Bose superfluid with the
  molecular scattering length $a_{\protect \mathrm {B}}=2a_s$.}\BibitemShut
  {Stop}%
\bibitem [{Note2()}]{note2}%
  \BibitemOpen
  \bibinfo {note} {We note that the Hartree self-energy in Fig.~\ref {fig4}(c)
  does not appear in $G_{\protect \rm B}$ at $T_{\protect \rm c}$. This is
  because it is canceled out by the Bose chemical potential, to give gapless
  Bose excitations at the superfluid phase transition.}\BibitemShut {Stop}%
\end{thebibliography}
\par
\end{document}